\documentclass[journal,10pt]{IEEEtran}
\usepackage{array}
\usepackage[caption=false,font=normalsize,labelfont=sf,textfont=sf]{subfig}
\usepackage{stfloats}
\usepackage{url}
\usepackage{cite}

\usepackage{adjustbox}
\usepackage{amsmath,amssymb,amsfonts}
\usepackage{graphicx}
\usepackage{tikz}
\usepackage{scalefnt}
\usepackage{fix-cm} 
\usepackage{multirow}
\usepackage{rotating}
\usepackage{textcomp}
\usepackage{mathrsfs}
\usepackage{xcolor}
\usepackage{blindtext}
\usepackage{pgfplots}
\usepgfplotslibrary{groupplots}
\usepackage{makecell}
\usepackage{algorithm}
\usepackage{algpseudocode}

\usetikzlibrary{spy}
\usepackage{authblk}
\usepackage{verbatim}

\usetikzlibrary{arrows.meta}
\usetikzlibrary{decorations.markings}

\usetikzlibrary{external}
\usepackage{amsthm}
\newtheorem{theorem}{Theorem}
\newtheorem{lemma}{Lemma}
\newtheorem{proposition}{Proposition}
\newtheorem{definition}{Definition}
\DeclareMathOperator*{\argmax}{arg\,max}

\newcommand\raisepunct[1]{\,\mathpunct{\raisebox{0.5ex}{#1}}}

\pgfplotsset{compat=1.16}

\begin{document}

\title{A matrix-inverse-free implementation of the MU-MIMO WMMSE beamforming algorithm}

\author{Lissy Pellaco and Joakim Jaldén
        % <-this % stops a space
\thanks{ This work was supported by the European Research Council project AGNOSTIC (742648). Lissy Pellaco and Joakim Jald\'en are with
    the Division of Information Science and Engineering, School of Electrical Engineering and Computer Science, KTH Royal Institute of Technology, SE-100 44 Stockholm, Sweden. (e-mail: pellaco@kth.se and jalden@kth.se).}
    }

\maketitle

\begin{abstract}
The WMMSE beamforming algorithm is a popular approach to address the NP-hard weighted sum rate (WSR) maximization beamforming problem. Although it efficiently finds a local optimum, it requires matrix inverses, eigendecompositions, and bisection searches, operations that are problematic for real-time implementation. 

In our previous work, we considered the MU-MISO case and effectively replaced such operations by resorting to a first-order method. Here, we consider the more general and challenging MU-MIMO case. Our earlier approach does not generalize to this scenario and cannot be applied to replace all the hard-to-parallelize operations that appear in the MU-MIMO case. Thus, we propose to leverage a reformulation of the auxiliary WMMSE function given by Hu \textit{et al}. By applying gradient descent and Schulz iterations, we formulate the first variant of the WMMSE algorithm applicable to the MU-MIMO case that is
free from matrix inverses and other serial operations and hence amenable to both real-time implementation and deep unfolding. 
From a theoretical viewpoint, we establish its convergence to a stationary point of the WSR maximization problem. From a practical viewpoint, we show that in a deep-unfolding-based implementation, the matrix-inverse-free WMMSE algorithm attains, within a fixed number of iterations, a WSR comparable to the original WMMSE algorithm truncated to the same number of iterations, yet with significant implementation advantages in terms of parallelizability and real-time execution.
\end{abstract}


\section{Introduction}
The choice of beamformers plays a crucial role %and is in fact one of the key enablers
for 5G and beyond networks~\cite{Chataut2020}, where weighted sum rate (WSR) maximization through interference management is often a primary
objective.
However, computing the optimal transmit beamformers for multi-user multiple-input multiple-output (MU-MIMO) downlink channels %with a total transmit power constraint
entails solving a non-convex NP-hard optimization problem~\cite{Luo2008,Liu2011}. Suboptimal solutions exist~\cite{Nguyen2014,Shi2011,Christensen2008,Ghauch2017,Kaleva2012,Cirik2015,Aquilina2017}, but their computational complexity and latency make them impractical for actual deployment. In particular, the WMMSE algorithm~\cite{Shi2011}, while capable of achieving a satisfactory WSR, is not a viable solution for new and future applications that demand high efficiency and extremely low latency.
Its iterative nature coupled with computationally intensive and hard-to-parallelize operations, such as eigendecompositions, bisection searches and matrix inverses, make the overall algorithm difficult to implement on real-time systems~\cite{He2022}. This has driven researchers to develop low-complexity and low-latency variants of the WMMSE algorithm, or surrogates thereof, that are appropriate for modern hardware platforms which rely on parallel computations for efficiency.
Further, the importance to replace matrix inverses with parellizable operations is also recognized by~\cite{Zhang2021}. %, where the authors use a deep neural network to substitute a bisection method that would otherwise require matrix inverses., although focused on hybrid precoding,

Next to black-box-based approaches~\cite{Sun2018,Xia2019,Huang2019,Huang2019b,Lu2020,Kim2020,Liang2020,Sun2021,Zhang2022}, in which the cumbersome operations of matrix inverse, eigendecomposition, and bisection search are replaced by deep neural networks trained to either approximate the output of the WMMSE algorithm or to directly maximize the WSR, deep-unfolding-based approaches have recently emerged~\cite{Chowdhury2021,Hu2021,Bao2021,Pellaco2021}.
While black-box approaches can be efficiently executed in parallel, they suffer from poor interpretability, entail the problem of architecture selection, and typically have a large number of parameters which translates to intensive training.
On the contrary, deep-unfolding-based approaches surpass the issue of architecture selection by constraining the structure of neural networks to replicate layer-by-layer the structure of the WMMSE algorithm, make the networks more easily interpretable and predictable, and generally lead to a reduced-dimensionality trainable space~\cite{Jagannath2021,Balatsouka2019,Monga2021}.
%%%%%%%%%%%%%%%%%%%%%%%%%%%%%%%

In~\cite{Chowdhury2021}, the authors built a network architecture that mimics a finite number of trainable WMMSE algorithm iterations in which learning modules are inserted to grant more flexibility.
%Despite the satisfactory performance and the acceleration in convergence, 
Despite the acceleration in convergence, the proposed approach still heavily relies on matrix inverses, which appear in each algorithm iteration. %and which are computationally prohibiting for real-time applications.
In~\cite{Hu2021, Bao2021,Pellaco2021}, the authors strived to remove matrix inverses within the WMMSE algorithm iterations to make the unfolded algorithm more amenable to real-time implementation. In~\cite{Hu2021}, the authors approximate matrix inversions with a trainable module structured as the first-order Taylor expansion of the matrix inverse operation and introduce the IAIDNN algorithm. However, this affects the achieved WSR and to compensate for such performance degradation, the authors i) include an explicit inverse in the initialization and in the final step of the algorithm and ii) restrict the use of such matrix inverse surrogate to the lightly loaded scenario, in which the first-order Taylor expansion attains a better approximation. In~\cite{Bao2021}, the authors adopt the same matrix approximation as~\cite{Hu2021}, but limit the computation of the explicit inverse to the initialization step. However, the resulting algorithm shows a non-monotic (oscillatory) behaviour in the WSR over the number of algorithm iterations. In~\cite{Pellaco2021}, we propose a matrix-inverse-free formulation of the WMMSE algorithm by replacing the method of Lagrange multipliers, which leads to a matrix inverse, with a first-order method.
However, this strategy is constrained to the multi-user multiple-input \emph{single-output} (MU-MISO) case and generally cannot be applied to circumvent all the matrix inverses that appear in the WMMSE algorithm in the MU-MIMO scenario.

As such, the state-of-the-art approaches present limitations, and a formulation of the WMMSE algorithm without any matrix inverse for the MU-MIMO scenario needs to be properly addressed. In this paper, we tackle this problem in a rigorous way and propose the first variant of the WMMSE algorithm completely free from matrix inverses (and other non-parallelizable operations) in the general MU-MIMO case. We refer to this algorithm as \textit{matrix-inverse-free WMMSE} algorithm and we detail our contributions below:
\begin{itemize}
    \item By resorting to the gradient descent and to the (Newton) Schulz iterative approach~\cite{Schulz1933,Householder1964} we effectively replace all matrix inverses in the original WMMSE algorithm with operations that can leverage parallel implementation and  hence support real-time implementation.
    
    \item We formally establish that the matrix-inverse-free WMMSE algorithm i) maintains the monotonicity of the auxiliary objective introduced by Shi \textit{et al.} in the original WMMSE algorithm and ii) converges to a stationary point of the \mbox{NP-hard} non-convex WSR maximization problem.

    \item To exemplify the relevance of our theoretical contribution, we unfold a finite number of iterations of the matrix-inverse-free WMMSE algorithm and insert trainable parameters to boost the achievable performance within the fixed computational complexity. To this end, we propose a more flexible variant of the matrix-inverse-free WMMSE algorithm, which is even more suitable to leverage the benefits of deep-unfolding-based implementation albeit missing some of the theoretical guarantees.
    
    \item  We show, by means of numerical results, that the unfolded matrix-inverse-free WMMSE network i) attains a comparable and often better performance than the WMMSE algorithm (truncated to the same number of iterations) while using only parallelizable operations and ii) surpasses IAIDNN~\cite{Hu2021} in the fully loaded scenario, while it performs on par in the lightly loaded scenario. The code to reproduce all the numerical results is available at~\cite{GITHUB}.
\end{itemize}

%%%%%%%%%%%%%%%%%%%%%%%

\subsection{Notation}
%We adopt the following notation.
We denote matrices and column vectors by bold uppercase and bold lowercase letters, respectively. $\boldsymbol{0}$ and $\boldsymbol{I}$ are the zero and identity matrix whose dimensions are made clear from context. We denote the transpose and the Hermitian transpose of a vector or matrix by $(\cdot)^T$ and $(\cdot)^H$, respectively, the \mbox{2-norm} and the Frobenius norm of a matrix by $\left\lVert \cdot \right\rVert_2$ and $\left\lVert \cdot \right\rVert_F$, respectively, and we adopt $(\cdot)^{-1}$, $\mathtt{Tr}(\cdot)$, $\mathtt{diag}(\cdot)$, $\sigma_i(\cdot)$, and $\lambda_i(\cdot)$ to indicate the inverse, the trace, the diagonal elements, the $i^{th}$ singular value, and the $i^{th}$ eigenvalue of a matrix, respectively.
For Hermitian matrices $\boldsymbol{X}$ and $\boldsymbol{Y}$, we use $\boldsymbol{X} \succeq \boldsymbol{Y}$ to indicate that $\boldsymbol{X} - \boldsymbol{Y}$ is positive semidefinite (PSD).
We indicate with $\boldsymbol{X}_{ij}$ the element of matrix $\boldsymbol{X}$ in row $i$ and column $j$.
For $\boldsymbol{X} \succeq 0 $, we denote with $\boldsymbol{X}^{\frac{1}{2}}$ the PSD Hermitian square root for which $\boldsymbol{X} = \boldsymbol{X}^{\frac{1}{2}} \boldsymbol{X}^{\frac{1}{2}}$. %We denote with $\boldsymbol{x}^{\dagger}$ the vector representation of matrix $\boldsymbol{X}$ given the concatenation of the transposed rows.
We denote the $M$-dimensional complex space by  $\mathbb{C}^{M}$.
We denote the expected value of $x$ computed with respect to the probability distribution of $y$ by $\mathbb{E}_y(x)$ and the gradient of $f$ with respect to $x$ by $\nabla_x f$. When clear from the context we omit subscripts.

\section{System Model}
We consider a multi-user multiple-input multiple-output (MU-MIMO) interference downlink channel. The base station (which has perfect channel knowledge) is equipped with $M$ antennas and sends $d$ independent data streams to $K$ users, each equipped with  $N$ antennas.
The base station uses beamformer $\boldsymbol{V}_k \in  \mathbb{C}^{M\times d}$ to transmit information to user $k$. The received signal at user $k$ is 
\begin{equation}
 \boldsymbol{y}_k = \boldsymbol{H}_k \boldsymbol{V}_k \boldsymbol{s}_k  + \sum_{m \neq k }^{K}{\boldsymbol{H}_k \boldsymbol{V}_m \boldsymbol{s}_m} + \boldsymbol{n}_k,
\end{equation}
where $\boldsymbol{H}_k \in \mathbb{C}^{N\times M}$ is the channel between the base station and user $k$, $\boldsymbol{s}_k \in \mathbb{C}^{d}$ is the signal to user $k$, and $n_k \in \mathbb{C}^{N}$ is (rotationally symmetric) additive white Gaussian noise experienced by user $k$, with zero mean and variance $\sigma^2$.
%distribution $\mathcal{CN}(0,\,\boldsymbol{I}\sigma^{2})$.
We assume $ \mathbb{E}[\boldsymbol{s}_k \boldsymbol{s}_k^H]= \boldsymbol{I}$. User $k$ adopts beamformer $\boldsymbol{U}_k  \in \mathbb{C}^{N\times d}$ to estimate the received signal
$\hat{\boldsymbol{s}}_k = \boldsymbol{U}^H_k \boldsymbol{y}_k$
and \mbox{the mean square error (MSE) is} 
\begin{equation}\label{eq:E}
\begin{aligned}
\widetilde{\boldsymbol{E}}_k & \triangleq \mathbb{E}_{\boldsymbol{s,n}}{\big[(\hat{\boldsymbol{s}}_k - \boldsymbol{s}_k)(\hat{\boldsymbol{s}}_k - \boldsymbol{s}_k)^H\big]} = \\&(\boldsymbol{I} - \boldsymbol{U}^{H}_k \boldsymbol{H}_k \boldsymbol{V}_k) (\boldsymbol{I} - \boldsymbol{U}^{H}_k \boldsymbol{H}_k \boldsymbol{V}_k)^H + \\&\sum_{m\neq k}{\boldsymbol{U}^H_k \boldsymbol{H}_k \boldsymbol{V}_m  \boldsymbol{V}^H_m \boldsymbol{H}^H_k \boldsymbol{U}_k }  + \sigma^2 \boldsymbol{U}^H_k \boldsymbol{U}_k \, .
\end{aligned}
\end{equation}
We aim to solve 
\begin{subequations}\label{eq:WSR_maximization}
\begin{alignat}{2}
\max_{\{\boldsymbol{V}_k\}_{k \in \mathcal{K}}} \quad & \sum_{k = 1}^{K}\alpha_k R_k \label{eq:WSR_maximization_cost_function}\\
\textrm{s.t.} \quad  & \sum_{k=1}^{K}{\mathtt{Tr}(\boldsymbol{V}_k\boldsymbol{V}_k^H)\leq P},
\end{alignat}
\end{subequations}
where $\mathcal{K} = \{1,2,\cdots,K \}$, $\alpha_k$ is the priority of user $k$ (assumed to be known), $R_k$ is the rate of user $k$ defined in~(\ref{eq:R}), and $P$ is the maximum transmit power. The received power is given both by the channel gain and the transmit power. Thus, we assume that $\lVert \boldsymbol{H}_k \rVert^2_2 \leq \kappa$ $\forall k$, in which case the received power is upper bounded by $P\kappa $, which will later be used in the proofs.

Unfortunately, problem~(\ref{eq:WSR_maximization}), is non-convex and NP-hard~\cite{Liu2011,Luo2008}. However, the popular and, by now, well-known WMMSE algorithm~\cite{Shi2011} finds a stationary point of~(\ref{eq:WSR_maximization}) by elegant use of an equivalent reformulation of~(\ref{eq:WSR_maximization}), as explained below. 
\begin{figure*}
\begin{equation}\label{eq:R}
    R_k \triangleq \log \det \Big( \boldsymbol{I} + \boldsymbol{H}_k \boldsymbol{V}_k \boldsymbol{V}_k^H \boldsymbol{H}_k^H \big(  \sum_{m \neq k}{\boldsymbol{H}_k \boldsymbol{V}_m \boldsymbol{V}_m^H \boldsymbol{H}_k^H} + \sigma^2 \boldsymbol{I} \big)^{-1} \Big).
\end{equation}
\end{figure*}

\section{The original WMMSE algorithm}
In~\cite{Shi2011}, Shi \textit{et al.} proposed to address a more tractable reformulation of problem~(\ref{eq:WSR_maximization}) given by
\begin{subequations}\label{eq:equivalent_reformulation}
\begin{alignat}{2}
\min_{\{\boldsymbol{U}_k,\boldsymbol{W}_k,\boldsymbol{V}_k\}_{k \in \mathcal{K}}} \quad & \sum_{k = 1}^{K}\alpha_k \big\{  \mathtt{Tr}(\boldsymbol{W}_k\widetilde{\boldsymbol{E}}_k) - \log\det(\boldsymbol{W}_k)\big\} \label{eq:cost_function} \\
\textrm{s.t.} \quad  & \sum_{k=1}^{K}{\mathtt{Tr}(\boldsymbol{V}_k\boldsymbol{V}_k^H)\leq P}, \label{eq:power_const}
\end{alignat}
\end{subequations}
where $\widetilde{\boldsymbol{E}}_k = \widetilde{\boldsymbol{E}}_k(\boldsymbol{U}_k,\{\boldsymbol{V}_k\}_{k \in \mathcal{K}})$ is the MSE matrix of user $k$ defined in~(\ref{eq:E}) and $\boldsymbol{W}_k \succeq 0 \in \mathbb{C}^{d \times d}$ is a weight matrix for user $k$. The advantage of addressing~(\ref{eq:equivalent_reformulation}) instead of~(\ref{eq:WSR_maximization}) is that the former is convex in each optimization variable $\{{\boldsymbol{U}_k}\}_{k \in \mathcal{K}}$, $\{{\boldsymbol{V}_k}\}_{k \in \mathcal{K}}$, and $\{{\boldsymbol{W}_k}\}_{k \in \mathcal{K}}$. It naturally remains non-convex in the joint space, owing to the NP-hard nature of the original problem. However, by convexity in each variable it is amenable to block coordinate descent and it is guaranteed to converge to a stationary point of problem~(\ref{eq:WSR_maximization})~\cite[Theorem 3]{Shi2011}.
This said, the resulting iterative algorithm~\cite[Table 1]{Shi2011} in its original form involves matrix inverses, eigendecompositions, and bisection searches at each iteration, operations that are computationally intensive and that significantly hinder real-time implementation.

A key observation made in~\cite{Hu2021} is that the power constraint in
problem~(\ref{eq:equivalent_reformulation}) can be dropped provided that i) in the cost function~(\ref{eq:cost_function}) $\widetilde{\boldsymbol{E}}$ is replaced by  $\boldsymbol{E}$ defined below in~(\ref{eq:E_tilde}) and \mbox{ii) that} the new unconstrained optimization problem is followed by a scaling operation to satisfy the power constraint~(\ref{eq:power_const}) with equality again. As argued in~\cite{Hu2021}, this equivalently leads to a stationary point of~(\ref{eq:equivalent_reformulation}).
In other words, instead of~(\ref{eq:equivalent_reformulation}) we can address the following unconstrained optimization problem
\begin{equation}\label{eq:equivalent_reformulation_tilde}
\min_{\{\boldsymbol{U}_k,\boldsymbol{W}_k,\boldsymbol{V}_k\}_{k \in \mathcal{K}}} \sum_{k = 1}^{K}\alpha_k \big\{  \mathtt{Tr}(\boldsymbol{W}_k\boldsymbol{E}_{k}) - \log\det(\boldsymbol{W}_k)\big\}
\end{equation}
where
%\begin{equation}
\begin{align}\label{eq:E_tilde}
    &\boldsymbol{E}_{k} \triangleq (\boldsymbol{I} - \boldsymbol{U}_k^H\boldsymbol{H}_k\boldsymbol{V}_k)(\boldsymbol{I} - \boldsymbol{U}_k^H\boldsymbol{H}_k\boldsymbol{V}_k)^H + \\&\!\!\sum_{m\neq k}^{K}{\!\boldsymbol{U}^H_k \boldsymbol{H}_k \boldsymbol{V}_m  \boldsymbol{V}^H_m \boldsymbol{H}^H_k \boldsymbol{U}_k } \! + \frac{\sum_{m=1}^{K}{\mathtt{Tr}(\boldsymbol{V}_m\boldsymbol{V}^H_m)}}{P}\sigma^2 \boldsymbol{U}^H_k \boldsymbol{U}_k \nonumber
\end{align}
%\end{equation}
followed by the scaling operation
\begin{equation}\label{eq:scaling_V}
 \boldsymbol{V}_k = \boldsymbol{V}_k\sqrt{\frac{P}{\sum_{m=1}^{K}{\mathtt{Tr}(\boldsymbol{V}_m \boldsymbol{V}^H_m)}}} \; \; k = 1,\cdots,K. 
\end{equation}
Like~(\ref{eq:equivalent_reformulation}), problem~(\ref{eq:equivalent_reformulation_tilde}) is non-convex in the joint space, but convex in each optimization variable. Thus, it is amenable to block coordinate descent. By sequentially optimizing over  $\{{\boldsymbol{U}_k}\}_{k \in \mathcal{K}}$,  $\{{\boldsymbol{W}_k}\}_{k \in \mathcal{K}}$, and  $\{{\boldsymbol{V}_k}\}_{k \in \mathcal{K}}$, it leads to iterative update equations that represent the optimal closed-form solution of each sub-problem, i.e.,
\begin{subequations}\label{eq:wmmse}
\begin{alignat}{3}
&\boldsymbol{U}_k = \boldsymbol{Q}_k^{-1} \boldsymbol{H}_k \boldsymbol{V}_k  &\mathrm{for}\;k = 1,\cdots,K \label{eq:u_update}\\
&\boldsymbol{W}_k = \boldsymbol{E}^{-1}_{k} = \big( \boldsymbol{I} - \boldsymbol{U}_k^H \boldsymbol{H}_k \boldsymbol{V}_k\big)^{-1} &\mathrm{for}\; k = 1,\cdots,K \label{eq:w_update}\\
&\boldsymbol{V}_k = \alpha_k\boldsymbol{R}^{-1} \boldsymbol{H}^H_k \boldsymbol{U}_k \boldsymbol{W}_k &\mathrm{for}\; k = 1,\cdots,K \label{eq:v_update}
\end{alignat}
\end{subequations}
where 
\begin{equation}\label{eq:Q}
    \boldsymbol{Q}_k = \frac{\sigma^2}{P}\sum_{m=1}^{K}{\mathtt{Tr}(\boldsymbol{V}_m \boldsymbol{V}_m)\boldsymbol{I}} + \sum_{m=1}^{K}{\boldsymbol{H}_k\boldsymbol{V}_m\boldsymbol{V}^H_m\boldsymbol{H}^H_k}
\end{equation}
and
\begin{equation}\label{eq:R_matrix}
    \boldsymbol{R} = \sum_{k = 1}^{K}\{\frac{\sigma^2}{P}{\mathtt{Tr}(\alpha_k \boldsymbol{U}_k \boldsymbol{W}_k \boldsymbol{U}^H_k) \boldsymbol{I}} + {\alpha_k \boldsymbol{H}^H_k \boldsymbol{U}_k \boldsymbol{W}_k \boldsymbol{U}^H_k \boldsymbol{H}_k}\}.
\end{equation}
Although it is a close variant of the original formulation proposed by Shi \textit{et al.}, we refer to it as the \textit{WMMSE algorithm} in the rest of the paper.

The key advantage of working with problem~(\ref{eq:equivalent_reformulation_tilde}) instead of~(\ref{eq:equivalent_reformulation}) is that the resulting algorithm does not involve any eigendecompositions and bisection searches. While this leads to an improvement in the overall computational complexity, the WMMSE algorithm still relies on matrix inverses in each algorithm iteration, as can be seen in~(\ref{eq:wmmse}).
\section{The matrix-inverse-free WMMSE algorithm}
Our main goal is to replace the complex matrix inverses that appear in the update equations of\footnote{\label{note1}In the rest of the paper, we use $\boldsymbol{X}$ as shorthand notation for $\{ \boldsymbol{X}_k\}_{k\in \mathcal{K}}$} $\boldsymbol{U}$~(\ref{eq:u_update}), $\boldsymbol{W}$~(\ref{eq:w_update}), and $\boldsymbol{V}$~(\ref{eq:v_update}) with more efficient operations that can leverage parallel implementation. 
In our previous paper~\cite{Pellaco2021}, which addresses the MU-MISO scenario, we proposed to circumvent the matrix inverse in $\boldsymbol{V}$ by resorting to projected gradient descent steps. In the MISO case only the update of $\boldsymbol{V}$ involves matrix inverse operations because $\boldsymbol{U}$ and $\boldsymbol{W}$ are scalars. Conversely, in the MU-MIMO scenario, all the three variables ($\boldsymbol{U}$, $\boldsymbol{W}$, and $\boldsymbol{V}$) are matrices and all the updates involve matrix inverse operations. For $\boldsymbol{V}$, we can apply the gradient descent (GD) approach as in~\cite{Pellaco2021} with the exception that the reformulation proposed by~\cite{Hu2021} obviates the need for the projection.
For $\boldsymbol{U}$, we can adopt the same approach as for $\boldsymbol{V}$, but for $\boldsymbol{W}$ we cannot. The update of $\boldsymbol{W}$ is problematic. It is imperative that $\boldsymbol{W}$ is a Hermitian positive semidefinite matrix throughout the entire algorithm. If it is not, cost functions~(\ref{eq:equivalent_reformulation_tilde}) and~(\ref{eq:cost_function}) lose meaning and the equivalence between problems~(\ref{eq:equivalent_reformulation}) and~(\ref{eq:WSR_maximization}) does not hold anymore. The gradient approach is not a viable option. First, it does not easily guarantee the positive semidefiniteness of $\boldsymbol{W}$.  Second, the gradient of~(\ref{eq:equivalent_reformulation_tilde}) with respect to $\boldsymbol{W}$ itself involves a matrix inverse and this naturally constitutes a major obstacle. Therefore, we are forced to abandon the gradient descent approach and look for an alternative.

To this end, we observe that the optimal update of $\boldsymbol{W}$ is given by $\boldsymbol{E}^{-1}$, yet we cannot compute  the matrix inverse operation directly. Instead, we propose to adopt the Schulz iterative approach, a previously recognized technique in the literature~\cite{Schulz1933,Householder1964} to iteratively compute matrix inverses. 
In particular, the Schulz iterative approach (specified below in~(\ref{eq:Schulz})) is guaranteed to converge to the true inverse for any initial point $\boldsymbol{W}$ satisfying
\begin{equation}\label{eq:Schulz_init}
 \rVert \boldsymbol{I} - {\boldsymbol{E}} \boldsymbol{W} \lVert_2 \;< 1. 
\end{equation}
Obviously, it is neither practical nor desirable to check this condition via a singular value decomposition before updating $\boldsymbol{W}$ in each algorithm iteration. 
However, intuitively, we can assume that if $\boldsymbol{W}$ is sufficiently close to $\boldsymbol{E}^{-1}$ in the first algorithm iteration and if the GD step sizes in the subsequent updates of $\boldsymbol{V}$ and $\boldsymbol{U}$ are bounded such that the perturbation of $\boldsymbol{E}$ (and hence of its inverse) is limited, then the Schulz iterations will be able to keep up with a sufficiently good estimate of $\boldsymbol{E}^{-1}$ throughout all algorithm iterations. In Section~\ref{sec:Monotonicity and convergence proof}, we will formally prove
that this is indeed the case.

To be concrete, we define objective function~(\ref{eq:equivalent_reformulation_tilde}) as
$f(\boldsymbol{U},\boldsymbol{W},\boldsymbol{V})\triangleq \sum_{k = 1}^{K}\alpha_k \big\{  \mathtt{Tr}(\boldsymbol{W}_k{\boldsymbol{E}}_{k}) - \log\det(\boldsymbol{W}_k)\big\}
$ and we provide in the following the alternative matrix-inverse-free updates of $\boldsymbol{U}$, $\boldsymbol{W}$, and $\boldsymbol{V}$ that we propose.
\subsubsection{Update of $\boldsymbol{U}$} 
The cost function~(\ref{eq:equivalent_reformulation_tilde}) is quadratic (convex) in $\boldsymbol{U}$ and we approximately solve 
\begin{equation}\label{eq:equivalent_reformulation_tilde_u}
\min_{\boldsymbol{U}} \sum_{k = 1}^{K}\alpha_k \big\{  \mathtt{Tr}\big(\boldsymbol{W}_k{\boldsymbol{E}}_{k}(\boldsymbol{U}_k,\boldsymbol{V})\big) - \log\det(\boldsymbol{W}_k)\big\},
\end{equation}
by applying the GD iterative approach.
We take $J_u$ GD steps of size $\gamma_{u}$ along the negative direction of the gradient. A single GD step is given by% In other words, the alternative update of $\boldsymbol{U}$ is given by 
%Thus, the alternative update of $\boldsymbol{U}$ is given by
\begin{equation}\label{eq:GD_u}
    \boldsymbol{U}^{+}_k =  \boldsymbol{U}_k -\gamma_{u}\nabla_{{u}} f\big(\boldsymbol{U}_k\big) \; \mathrm{for} \; k =1,\cdots,K,
\end{equation}
where the superscript $+$ indicates the updated variable and $\nabla_{{u}} f$ is the gradient of~(\ref{eq:equivalent_reformulation_tilde_u}) with respect to $\boldsymbol{U}$, i.e.,
%\begin{equation}
\begin{align}\label{eq:gradient_u_expression}
    \nabla_{{u}} f\big(\boldsymbol{U}_k\big) &= 2\alpha_k \big(-\boldsymbol{H}_k \boldsymbol{V}_k +   \sum_{m}^{}{\boldsymbol{H}_k \boldsymbol{V}_m  \boldsymbol{V}^H_m} \boldsymbol{H}^H_k \boldsymbol{U}_k\big)\boldsymbol{W}_k \nonumber \\ &+   \frac{2\alpha_k \sigma^2}{P}\sum_{n = 1}^{K}{\mathtt{Tr}( \boldsymbol{V}_n  \boldsymbol{V}^{H}_n )} \boldsymbol{U}_k \boldsymbol{W}_k.
\end{align}
%\end{equation}
%%%%%%%%%%%%%%%%%%%%%%%%%%%%%%%%%%%%%%%%%%%%
\subsubsection{Update of $\boldsymbol{W}$}
As mentioned before, the optimal update of $\boldsymbol{W}$ is given by the inverse of ${\boldsymbol{E}}$, an operation that we propose to compute approximately and implicitly by applying Schulz iterative approach~\cite{Schulz1933}. A single Schulz iteration is given by
\begin{equation}\label{eq:Schulz}
    \boldsymbol{W}^{+}_k = \boldsymbol{W}_k(2\boldsymbol{I} - {\boldsymbol{E}}_k\boldsymbol{W}_k) \; \; \mathrm{for}\; k=1,\cdots,K,
\end{equation}
where the superscript $+$ indicates the updated variable. Most importantly, each iteration ensures the positive semidefiniteness of $\boldsymbol{W}^+$, provided that $\boldsymbol{W}$ is positive semidefinite itself, as later established in Lemma~\ref{lem:combined_Schulz_update}.
%%%%%%%%%%%%%%%%%%%%%%%%%%%%%%%%%%%%%%%%%%%%%%%%%%%%
\subsubsection{Update of $\boldsymbol{V}$} 
The cost function~(\ref{eq:equivalent_reformulation_tilde}) is quadratic (convex) in $\boldsymbol{V}$, thus we can apply the GD approach and 
approximate
\begin{equation}\label{eq:equivalent_reformulation_tilde_v}
\min_{\boldsymbol{V}} \sum_{k = 1}^{K}\alpha_k \big\{  \mathtt{Tr}\big(\boldsymbol{W}_k{\boldsymbol{E}}_{k}(\boldsymbol{U}_k,\boldsymbol{V})\big) - \log\det(\boldsymbol{W}_k)\big\},
\end{equation}
by taking $J_v$ GD steps of size $\gamma_v$. A single step is given by
\begin{equation}\label{eq:GD_v}
    \boldsymbol{V}^{+}_k =  \boldsymbol{V}_k -\gamma_{v}\nabla_{{v}} f\big(\boldsymbol{V}_k\big) \; \mathrm{for} \; k =1,\cdots,K,
\end{equation}
where the superscript $+$ indicates the updated variable and $\nabla_{{v}} f$ is the gradient of~(\ref{eq:equivalent_reformulation_tilde_v}) with respect to $\boldsymbol{V}$, i.e.,
%\begin{equation}
\begin{align}\label{eq:gradient_v_expression}
    &\nabla f_{{v}}\big(\boldsymbol{V}_k\big) = \sum_{m=1}^{K}{2\alpha_m \boldsymbol{H}^H_m \boldsymbol{U}_m \boldsymbol{W}_m \boldsymbol{U}^H_m  \boldsymbol{H}_m \boldsymbol{V}_k} \\ &-2\alpha_k\boldsymbol{H}^H_k \boldsymbol{U}_k \boldsymbol{W}_k +  2 \boldsymbol{V}_k  \frac{\sigma^2}{P}\sum_{m = 1}^{K}{\alpha_m \mathtt{Tr}( \boldsymbol{W}_m  \boldsymbol{U}^{H}_m  \boldsymbol{U}_m )}.\nonumber
\end{align}
%\end{equation}
After each GD step, we normalize $\boldsymbol{V}$ such that the power constraint is met with equality and, in order to avoid perturbing the cost function~(\ref{eq:equivalent_reformulation_tilde}), we normalize $\boldsymbol{U}$ by the reciprocal scaling, i.e., 
$\boldsymbol{V} = \beta \boldsymbol{V}$  and $\boldsymbol{U} = \frac{1}{\beta} \boldsymbol{U}$, where $\beta = \sqrt{\frac{P}{\sum_{m}{\mathtt{Tr}(\boldsymbol{V}_m \boldsymbol{V}^H_m)}}}$, so that $\boldsymbol{E}$ remains constant.
Algorithm~\ref{alg:matrix_inverse_free_WMMSE} summarizes the overall algorithm. 

\begin{algorithm}
\caption{Matrix-inverse-free WMMSE algorithm}\label{alg:matrix_inverse_free_WMMSE}
\begin{algorithmic}
%\State{Set $\gamma_u \leq \min{(\gamma^*_u,\nu_u)} $ and  $\gamma_v \leq \min{(\gamma^*_v,\nu_v)}$}
\State {Initialize $\boldsymbol{V}$ such that $\sum_{m=1}^{K}{\mathtt{Tr}(\boldsymbol{V}_m\boldsymbol{V}_m^H) = P}$}
\State {Initialize $\boldsymbol{U} = \boldsymbol{0}$}
\State {Initialize $\boldsymbol{W} = \boldsymbol{I}$}
\For{$l=1,\cdots,L$}
\For {$j=1,\cdots,J_u$}
\State{$\boldsymbol{U}_k \gets  \boldsymbol{U}_k -\gamma_{u}\nabla_{{u}} f\big(\boldsymbol{U}_k\big) \; \mathrm{for} \; k =1,\cdots,K$.}
\EndFor
\For {$j=1,\cdots,J_w$}
\State{$\boldsymbol{W}_k \gets \boldsymbol{W}_k(2\boldsymbol{I} - {\boldsymbol{E}}_k\boldsymbol{W}_k) \; \mathrm{for} \; k =1,\cdots,K$}
\EndFor
\For {$j=1,\cdots,J_v$}
\State{$\boldsymbol{V}_k \gets  \boldsymbol{V}_k -\gamma_{v}\nabla_{{v}} f\big(\boldsymbol{V}_k\big) \; \mathrm{for} \; k =1,\cdots,K$.}
\State{$\boldsymbol{V} \gets \beta \boldsymbol{V}$ and $\boldsymbol{U} \gets \frac{1}{\beta}\boldsymbol{U}$, $\beta = \sqrt{\frac{P}{\sum_{m}{\mathtt{Tr}(\boldsymbol{V}_m \boldsymbol{V}^H_m)}}}$}
\EndFor
\EndFor
\end{algorithmic}
\end{algorithm}

The matrix-inverse-free WMMSE algorithm does not fully solve the optimization problems in $\boldsymbol{U}$, $\boldsymbol{W}$, and $\boldsymbol{V}$. Instead, it approximately solves them by taking a finite number of GD step in $\boldsymbol{U}$ and $\boldsymbol{V}$ and by performing a finite number of Schulz iterations in $\boldsymbol{W}$. This is in contrast to the WMMSE algorithm, which optimally solves each individual optimization problem in $\boldsymbol{U}$, $\boldsymbol{W}$, and $\boldsymbol{V}$ and is hence naturally guaranteed to monotonically decrease~(\ref{eq:equivalent_reformulation_tilde}) and to converge to a stationary point of~(\ref{eq:WSR_maximization}).
Nevertheless, we can and will below establish that the matrix-inverse-free WMMSE algorithm also monotonically decreases~(\ref{eq:equivalent_reformulation_tilde}) and converges to a stationary point of~(\ref{eq:WSR_maximization}), provided that $\gamma_u > 0$ and $\gamma_v >0$ are appropriately chosen. Thus, unlike black-box or approximate implementations, the matrix-inverse-free implementation retains the most important properties of the original WMMSE algorithm.

\section{Monotonicity and convergence proof}\label{sec:Monotonicity and convergence proof}
This section is devoted to establishing monotonicity (Theorem~\ref{theo:monotonicity}) and convergence  (Theorem~\ref{theo:stationary_point}) of the matrix-inverse-free WMMSE algorithm provided that $\gamma_u > 0$ and $\gamma_v > 0$ are sufficiently small.

\begin{theorem}\label{theo:monotonicity}
The matrix-inverse-free WMMSE algorithm with \mbox{$ 0 < \gamma_u \leq \min(\frac{1}{L_u},\nu_u)$} and with \mbox{$ 0 < \gamma_v \leq \min(\frac{1}{L_v},\nu_v)$}, where \mbox{$L_u = \frac{2\bar{\alpha}\delta(P\kappa + \sigma^2)^2}{\sigma^2}\raisepunct{,}$} \mbox{$L_v=\frac{2\bar{\alpha}K\delta(P\kappa +d\sigma^2 )}{P\sigma^2}\raisepunct{,}$} where $\bar{\alpha} \geq \alpha_k \,\forall k$ and $\nu_u$ and $\nu_v$ are specified in~(\ref{eq:def_nu_u}) and in~(\ref{eq:def_nu_v}), respectively, generates a sequence of iterates with monotonically non-increasing cost function $f$. Moreover, monotonicity holds for each individual variable update step in the algorithm.
\end{theorem}
To see the rationale behind Theorem~\ref{theo:monotonicity}, consider the updates of $\boldsymbol{U}$ and $\boldsymbol{V}$. Both updates are based on the gradient descent approach, which, in case of convex optimization, is guaranteed to monotonically decrease the cost function if the cost function is $L$-smooth~\cite{Nesterov2014}. 
In Lemmas~\ref{lem:L_smooth_U} and~\ref{lem:L_smooth_V}, we derive the upper bounds of the $L$-smooth constants of $f$ as a function of $\boldsymbol{U}$ and $\boldsymbol{V}$, i.e., $L_u$ and $L_v$, and we set $ \gamma_u \leq \frac{1}{L_u}$ and $ \gamma_v \leq  \frac{1}{L_v}$ to ensure monotonicity. %a monotonic decrement in $f$ over the $\boldsymbol{U}$ and the $\boldsymbol{V}$ update.

%Consider now the update of $\boldsymbol{W}$. It has
The Schulz update of $\boldsymbol{W}$ will later be shown to maintain positive semidefiniteness% of $\boldsymbol{W}$
, but it is more difficult to ensure monotonicity. However, Lemma~\ref{lem:combined_Schulz_update} will establish that if 
\begin{equation}\label{eq:condition_eig_EW_delta}
    0 \leq %\lambda_{i}(\boldsymbol{E}_k\boldsymbol{W}_k)
    \lambda_i(\boldsymbol{E}_k\boldsymbol{W}_k) =
    \sigma_i(\boldsymbol{W}_k^{\frac{1}{2}} \boldsymbol{E}_k \boldsymbol{W}_k^{\frac{1}{2}})
    \leq \delta \; \; \forall i,k
\end{equation}
holds, where $\delta \approx 1.68$ is the unique solution to $ x - x^2 - \log(2-x)  = 0$ in the range $1 < x < 2$, then the Schulz iterations will also give a monotonic decrement in $f$ (unless already at the optimum). Thus, it is sufficient to guarantee that condition~(\ref{eq:condition_eig_EW_delta}) is satisfied when executing the update.
Lemma~\ref{lem:combined_Schulz_update} also establishes that each Schulz iteration brings the eigenvalues of $\boldsymbol{E}_k\boldsymbol{W}_k$ into in the range $[0,1] \; \forall k$. %if~(\ref{eq:condition_eig_EW_delta}) is satisfied.
Thus, it is sufficient to ensure~(\ref{eq:condition_eig_EW_delta}) at the start of the $J_w$ iterations as $1 < \delta$. However, before taking the next $J_w$ Schulz steps, $\boldsymbol{E}$ is perturbed by the updates of $\boldsymbol{V}$ and $\boldsymbol{U}$ and this might in turn perturb the eigenvalues of $\boldsymbol{E}_k\boldsymbol{W}_k$ enough to violate condition~(\ref{eq:condition_eig_EW_delta}). Therefore, we need to limit the perturbation of $\boldsymbol{E}$ in order to ensure that condition~(\ref{eq:condition_eig_EW_delta}) is also satisfied after the updates of $\boldsymbol{V}$ and $\boldsymbol{U}$.
In Lemma~\ref{lem:step_size_bound_GD_UV}, we show that setting $ \gamma_u \leq \nu_u$ and $ \gamma_v \leq \nu_v$, where $\nu_u$ and $\nu_v$ are given in~(\ref{eq:def_nu_u}) and in~(\ref{eq:def_nu_v}), respectively, ensures that the perturbation of $\boldsymbol{E}$ caused by the updates of $\boldsymbol{V}$ and $\boldsymbol{U}$ does not violate~(\ref{eq:condition_eig_EW_delta}). The key observation used in the proof is that $\lVert \nabla_u f \rVert_2 $ and $\lVert \nabla_v f\rVert_2 $ are always universally bounded throughout the algorithm.

To summarize, the bounds $\gamma_u \leq \frac{1}{L_u}$ and $\gamma_v \leq \frac{1}{L_v}$ ensure that the updates of $\boldsymbol{U}$ and $\boldsymbol{V}$ monotonically decrease~(\ref{eq:equivalent_reformulation_tilde}) and the bounds $ \gamma_u \leq \nu_u$ and $\gamma_v \leq \nu_v$ ensure that if the initialization satisfies $\lambda_i(\boldsymbol{E}_k\boldsymbol{W}_k) \leq 1 \, \forall i,k$%~(\ref{eq:condition_eig_EW_delta})
, as in Algorithm~\ref{alg:matrix_inverse_free_WMMSE}, then~(\ref{eq:condition_eig_EW_delta}) will be satisfied throughout the entire algorithm and this guarantees that the update of $\boldsymbol{W}$ monotonically decreases~(\ref{eq:equivalent_reformulation_tilde}). We will restate this argument in Section~\ref{sec:monotonicity}, where we formally prove Theorem~\ref{theo:monotonicity} and provide the required lemmas. Before this, however, we provide a notationally convenient inner product and restate the definition of stationary point together with the formal convergence property of the matrix-inverse-free WMMSE algorithm proven in Section~\ref{sec:convergence}.
\begin{definition}\label{def:inner_product}
We define the inner product between complex matrices $\boldsymbol{X}$ and $\boldsymbol{Y}$, denoted by $\langle \boldsymbol{X},\boldsymbol{Y} \rangle$, as the Euclidean inner product between the real-valued vector representation of each matrix, obtained by concatenating the real part and the imaginary part of the transposed rows.
\end{definition}
\begin{proposition}[\hspace{-0.5pt}{\cite[Proposition 2.1.2]{Bertsekas99}}]\label{prop: optimality condition}
If $\overline{\boldsymbol{X}}$ is a local minimum of a (real-valued) continuously differentiable function $f$ over a convex set $\mathcal{X}$, then
\begin{equation}\label{def:stationary_point}
\langle \nabla f(\overline{\boldsymbol{X}}),\boldsymbol{X}-\overline{\boldsymbol{X}}\rangle \;\geq 0 \; \forall \boldsymbol{X} \in \mathcal{X},
\end{equation}
where $\langle \cdot,\cdot\rangle $ is given in Definition~\ref{def:inner_product}. Any $\overline{\boldsymbol{X}}$ satisfying~(\ref{def:stationary_point}) is referred to as a stationary point.
\end{proposition}
%Any $\overline{\boldsymbol{X}}$ satisfying~(\ref{def:stationary_point}) is referred to as a stationary point. 
\begin{theorem}\label{theo:stationary_point}
Any limit point $(\overline{\boldsymbol{U}},\overline{\boldsymbol{W}},\overline{\boldsymbol{V}})$ of the iterates generated by the matrix-inverse-free WMMSE algorithm under the conditions of Theorem~\ref{theo:monotonicity} as $L~\rightarrow~\infty$ is a stationary point of~(\ref{eq:equivalent_reformulation}) and the corresponding $\overline{\boldsymbol{V}}$ is a stationary point of~(\ref{eq:WSR_maximization}). Conversely, if $\overline{\boldsymbol{V}}$ is a stationary point of~(\ref{eq:WSR_maximization}), then $(\overline{\boldsymbol{U}},\overline{\boldsymbol{W}},\overline{\boldsymbol{V}})$ is a stationary point of~(\ref{eq:equivalent_reformulation}) for some $\overline{\boldsymbol{U}}$ and $\overline{\boldsymbol{W}}$.
\end{theorem}

%%%%%%%%%%%%%%%%%%%%%%%%%%%%%%%%%%%%%%%%%%%%%%%%%%%%%%%
\subsection{Monotonicity}\label{sec:monotonicity}
We begin by showing that the matrix-inverse-free
WMMSE algorithm yields a sequence $\{\boldsymbol{U},\boldsymbol{W},\boldsymbol{V}\}^l$ with
monotonically non-increasing cost function $f$ provided that $\gamma_u > 0$ and \mbox{$\gamma_v > 0$} are sufficiently small. We state and explain all the required lemmas, but we relegate their proofs to the Appendix.

We start by laying out some assumptions and notation.
Let $\bar{\alpha}$ be the largest user priority, i.e., 
$\bar{\alpha}=\max_k{\alpha_k}$, and recall that $\kappa$ bounds the largest singular value of $\boldsymbol{H}^H_k \boldsymbol{H}_k\; \forall k$, i.e., 
\begin{equation}\label{eq:kappa}
\sigma_{\mathrm{max}}(\boldsymbol{H}^H_k \boldsymbol{H}_k) \leq \kappa \quad \forall k,
\end{equation}
so that $\lVert \boldsymbol{H}_k \rVert_2 \leq \sqrt{\kappa} \; \forall k$.
We first assume that~(\ref{eq:condition_eig_EW_delta}) is satisfied throughout the entire algorithm and then show that this is the case if the initialization satisfies $\lambda_{\mathrm{max}}(\boldsymbol{E}_k\boldsymbol{W}_k) \leq 1 \, \forall k$.

We start by stating the following two lemmas, which establish fundamental properties that will be used throughout the proofs, as detailed in the Appendix. 
\begin{lemma}\label{lem:lower_bound_E}
For any $\boldsymbol{U}$ and $\boldsymbol{V}$ for which $\lVert \boldsymbol{V} \rVert^2_F \leq P$, $\lambda_{\mathrm{min}} (\boldsymbol{E}_k)\ge \frac{\sigma^2}{ P \kappa +\sigma^2} \; \forall k$, where $\boldsymbol{E}_k$ is defined in~(\ref{eq:E_tilde}). 
\end{lemma}

The following is a standard lemma from the literature but we give it here for completeness.
\begin{lemma}\label{lem:on_eigenvalues_of_EW}
If $\boldsymbol{E} \succeq 0$ and $\boldsymbol{W} \succeq 0$, then $\boldsymbol{EW}$ has real-valued and non-negative eigenvalues given by \mbox{$\lambda_i(\boldsymbol{E} \boldsymbol{W}) = \sigma_i(\boldsymbol{W}^{\frac{1}{2}} \boldsymbol{E} \boldsymbol{W}^{\frac{1}{2}})$} $\forall i $.
\end{lemma}

Lemma~\ref{lem:on_eigenvalues_of_EW} formally shows that the equality in~(\ref{eq:condition_eig_EW_delta}) is always true and it is equivalent to $\lambda_{\mathrm{max}}(\boldsymbol{E}_k\boldsymbol{W}_k) \leq \delta \;\; \forall k$.

We now consider the updates of $\boldsymbol{U}$ and $\boldsymbol{V}$ and we formalize the upper bounds $L_u$ (Lemma~\ref{lem:L_smooth_U}) and $L_v$ (Lemma~\ref{lem:L_smooth_V}) on the $L$-smooth constants of $f$ as a function of $\boldsymbol{U}$ and $\boldsymbol{V}$, respectively. These upper bounds are essential to guarantee monotonicity over the updates of $\boldsymbol{U}$ and $\boldsymbol{V}$. 
%We can now formalize $L_u$ (in Lemma~\ref{lem:L_smooth_U}) and $L_v$ (in Lemma~\ref{lem:L_smooth_V}).
\begin{lemma}\label{lem:L_smooth_U}
$f(\boldsymbol{U},\boldsymbol{W},\boldsymbol{V})$ is an $L$-smooth function over $\boldsymbol{U}$ with $L \leq L_u$, where \mbox{$L_u  = \frac{2\bar{\alpha}\delta(P\kappa + \sigma^2)^2}{\sigma^2}$} for any $\boldsymbol{W}$ and $\boldsymbol{V}$ generated by the matrix-inverse-free WMMSE algorithm. 
\end{lemma}

\begin{lemma}\label{lem:L_smooth_V}
$f(\boldsymbol{U},\boldsymbol{W},\boldsymbol{V})$ is an $L$-smooth function over $\boldsymbol{V}$ with $L\leq L_v$, where \mbox{$L_v=\frac{2\bar{\alpha}K\delta(P\kappa + d\sigma^2)}{P\sigma^2}$} for any $\boldsymbol{U}$ and $\boldsymbol{W}$ generated by the matrix-inverse-free WMMSE algorithm.
\end{lemma}

From the convexity of $f$ in each optimization variable, it follows from~\cite[Section 1.2.3]{Nesterov2014} that taking $J_u$ GD steps on $\boldsymbol{U}$ with $0 <\gamma_u\leq \frac{1}{L_u}$ and taking $J_v$ GD steps on $\boldsymbol{V}$ with $ 0 < \gamma_v \leq \frac{1}{L_v}$ yields a sequence with monotonically non-increasing cost function. We recall that the reciprocal scaling of $\boldsymbol{U}$ and $\boldsymbol{V}$ after each GD step on $\boldsymbol{V}$ does not perturb $\boldsymbol{E}$ (and consequently does not perturb $f$) and hence does not alter monotonicity.

We next consider the updates of $\boldsymbol{W}$ and show in Lemma~\ref{lem:combined_Schulz_update} that each Schulz iteration i) preserves monotonicity of $f$, ii)~grants the positive semidefiniteness of $\boldsymbol{W}$, and iii) brings the eigenvalues
of $\boldsymbol{E}_k \boldsymbol{W}_k$ in the range $[0, 1] \; \forall k$ if~(\ref{eq:condition_eig_EW_delta}) holds.
\begin{lemma}\label{lem:combined_Schulz_update}
If $\boldsymbol{W}_k \succeq 0$, $\boldsymbol{E}_k \succ 0$,  and $\lambda_{\mathrm{max}}(\boldsymbol{E}_k \boldsymbol{W}_k) \leq \delta$ $\forall k$, where $\delta \approx 1.68$ is the unique solution to $x-x^2 -\log (2-x) = 0$ in the range $1 < x < 2$, then $f(\boldsymbol{U},\boldsymbol{W}^+,\boldsymbol{V}) \leq f(\boldsymbol{U},\boldsymbol{W},\boldsymbol{V})$, $ \boldsymbol{W}^+_k \succeq 0$, and $\lambda_i(\boldsymbol{E}_k \boldsymbol{W}^+_k) \in [0,1] \;\forall i, k$, where $\boldsymbol{W}_k^+ = \boldsymbol{W}_k(2\boldsymbol{I} - \boldsymbol{E}_k\boldsymbol{W}_k) \; \forall k$. 
\end{lemma}

Note here that $\boldsymbol{E}\succ0$ is always true by Lemma~\ref{lem:lower_bound_E}. Although Lemma~\ref{lem:combined_Schulz_update} considers a single Schulz iteration, it can be easily extended to the case of an arbitrary finite number of Schulz iterations given that $\lambda_{\mathrm{max}}(\boldsymbol{E}_k \boldsymbol{W}^+_k) \leq 1 $ implies $\lambda_{\mathrm{max}}(\boldsymbol{E}_k \boldsymbol{W}^+_k) \leq \delta$, where $1 < \delta < 2$. Lemma~\ref{lem:combined_Schulz_update} can thus be recursively applied. If $\delta < \lambda_{\mathrm{max}}(\boldsymbol{E}_k \boldsymbol{W}_k) \leq 2$, then Schulz iterations still guarantee the positive semidefiniteness of $\boldsymbol{W}$ and bring the eigenvalues
of $\boldsymbol{E}_k \boldsymbol{W}_k$ in the range $[0, 1] \, \forall k$, although monotonicity is not guaranteed to hold. 

We finally consider the perturbation of $\boldsymbol{E}$~(\ref{eq:E_tilde}) given by the updates of $\boldsymbol{V}$ and $\boldsymbol{U}$ (Lemma~\ref{lem:step_size_bound_GD_UV}) and derive a bound on $\gamma_u$ and $\gamma_v$ to ensure that the perturbation of $\boldsymbol{E}$ is in turn bounded and does not violate~(\ref{eq:condition_eig_EW_delta}).

\begin{lemma}~\label{lem:step_size_bound_GD_UV}
If $ 0 < \gamma_u \leq \nu_u$ and $ 0 < \gamma_v \leq \nu_v$, where $\nu_u$ and $\nu_v$ are given in~(\ref{eq:def_nu_u}) and in~(\ref{eq:def_nu_v}), respectively, then $\lambda_{\mathrm{max}}(\boldsymbol{E}^+_k\boldsymbol{W}_k ) = \lambda_{\mathrm{max}}(\boldsymbol{W}_k \boldsymbol{E}^+_k) \leq \delta$, where $\boldsymbol{E}^+_k$ indicates $\boldsymbol{E}_k$~(\ref{eq:E_tilde}) evaluated after taking $J_v$ steps on $\boldsymbol{V}$ and $J_u$ steps on $\boldsymbol{U}$, as specified by Algorithm~\ref{alg:matrix_inverse_free_WMMSE}.
\end{lemma}

\begin{figure*}
\begin{equation}\label{eq:def_nu_u}
 \nu_u \triangleq \frac{-L_d + \sqrt{L_d^2 + 4L_c(\frac{\delta -1}{J_u +J_v})}}{2L_c} \raisepunct{,} \; \mathrm{where} \; L_c \triangleq \mu^2_u \frac{\delta (P\kappa + \sigma^2)^2}{\sigma^2}\raisepunct{,} \; L_d \triangleq 2\mu_u \delta( P\kappa + \sigma^2)\frac{ (\sqrt{P\kappa}+\sqrt{P\kappa + \sigma^2})}{\sigma^2}\raisepunct{,}
\end{equation}
and $\mu_u\triangleq  2\bar{\alpha}\delta \sqrt{ P\kappa + \sigma^2}(1 +\frac{P\kappa +\sqrt{P\kappa(P\kappa + \sigma^2)}}{\sigma^2})$.

\begin{equation}\label{eq:def_nu_v}
\nu_v \triangleq \frac{-L_b + \sqrt{L_b^2 + 4L_a(\frac{\delta -1}{J_u +J_v})}}{2L_a}\raisepunct{,} \; \mathrm{where} \; L_a \triangleq \frac{  K \mu^2_v\delta(\kappa + \frac{d \sigma^2}{P})}{\sigma^2}\raisepunct{,} \; L_b \triangleq \frac{2 K \sqrt{P} \mu_v\delta(\kappa + \frac{\sqrt{d}\sigma^2}{P})  + 2\mu_v \delta \sqrt{\kappa (P\kappa + \sigma^2) }}{\sigma^2}\raisepunct{,}
\end{equation}
and $\mu_v\triangleq 2 \bar{\alpha}\delta \frac{(KP \kappa  + \sqrt{P\kappa  ( P\kappa + \sigma^2) } + K d\sigma^2)}{\sqrt{P}\sigma^2 }\raisepunct{.}$
\end{figure*}

Therefore, we have shown that if the matrix-inverse-free WMMSE algorithm is initialized such that $\lambda_{\mathrm{max}}(\boldsymbol{E}_k \boldsymbol{W}_k) \leq 1 \, \forall k$, then~(\ref{eq:condition_eig_EW_delta}) will hold throughout the entire algorithm provided that $ \gamma_u \leq \min (\frac{1}{L_u},\nu_u)$ and $ \gamma_v \leq \min (\frac{1}{L_v},\nu_v)$ and this is key %property in order
to establish the monotonicity of the sequence of iterates. Note here also that the %It must be mentioned that the
initialization proposed in Algorithm~\ref{alg:matrix_inverse_free_WMMSE}, where $\boldsymbol{U} = \boldsymbol{0}$, implies
%, given by $\boldsymbol{W} = \boldsymbol{I}$ and $\boldsymbol{U} = \boldsymbol{0}$, yields
$\boldsymbol{E} = \boldsymbol{I}$ and $\lambda_i(\boldsymbol{E}_k \boldsymbol{W}_k) = 1 \; \forall i, k$, as $\boldsymbol{W} = \boldsymbol{I}$.
%and consequently satisfies condition~(\ref{eq:condition_eig_EW_delta}).

Now that we have in place all the required lemmas, we can provide the proof of Theorem~\ref{theo:monotonicity}.
%%%%%%%%%%%%%%%%%%%%%%%%%%%%%%%%%%%%%%%%%%%%%%%%%%%%%%%%%%%%%%%
\begin{proof}
For the purpose of the proof we do not consider the reciprocal scaling of $\boldsymbol{U}$ and $\boldsymbol{V}$, as it does not perturb $f$.
Let $\boldsymbol{Z}^{l}_{u}$ denote $(\boldsymbol{U}^{l+1},\boldsymbol{W}^{l},\boldsymbol{V}^{l})$, where $l$ is the iteration index, let $\boldsymbol{Z}^{l}_{w}$ denote $(\boldsymbol{U}^{l+1},\boldsymbol{W}^{l+1},\boldsymbol{V}^{l})$, and let $\boldsymbol{Z}^{l}_{v}$ denote $(\boldsymbol{U}^{l+1},\boldsymbol{W}^{l+1},\boldsymbol{V}^{l+1})$. Applying $J_u$ GD steps of size $ 0 < \gamma_u \leq \min(\frac{1}{L_u},\nu_u)$ to approximately solve
\begin{equation}
\min_{\boldsymbol{\xi}} f(\boldsymbol{\xi},\boldsymbol{W}^{l},\boldsymbol{V}^{l}),
\end{equation}
where $f(\boldsymbol{U},\boldsymbol{W}^{l},\boldsymbol{V}^{l})$ is $L$-smooth with $L\leq L_u$, as stated in Lemma~\ref{lem:L_smooth_U}, yields~\cite{Nesterov2014}
\begin{equation}
f(\boldsymbol{U}^{l},\boldsymbol{W}^{l},\boldsymbol{V}^{l}) \geq f(\boldsymbol{U}^{l+1},\boldsymbol{W}^{l},\boldsymbol{V}^{l}),
\end{equation}
namely
\begin{equation}\label{eq:inequality_monotonicity_1}
f(\boldsymbol{Z}^{l-1}_v)\geq f(\boldsymbol{Z}^l_u).
\end{equation}
%By using $ 0 < \gamma_u \leq \min(\frac{1}{L_u},\nu_u)$,
By Lemma~\ref{lem:step_size_bound_GD_UV} we have that $\lambda_{\mathrm{max}}(\boldsymbol{E}_k \boldsymbol{W}_k) \leq \delta$, which implies by Lemma~\ref{lem:combined_Schulz_update} that after $J_w$ Schulz iterations on $\boldsymbol{W}$ we have %that
\begin{equation}\label{eq:inequality_monotonicity_2}
f(\boldsymbol{Z}^{l}_u)\geq f(\boldsymbol{Z}^l_{w}).
\end{equation}
Finally, applying $J_v$ GD steps of size \mbox{$ 0 <\gamma_v \leq \min(\frac{1}{L_v},\nu_v)$} to approximately solve
\begin{equation}
\min_{\boldsymbol{\xi}} f(\boldsymbol{U}^{l+1},\boldsymbol{W}^{l+1},\boldsymbol{\xi})
\end{equation}
yields
\begin{equation}
f(\boldsymbol{U}^{l+1},\boldsymbol{W}^{l+1},\boldsymbol{V}^{l}) \geq f(\boldsymbol{U}^{l+1},\boldsymbol{W}^{l+1},\boldsymbol{V}^{l+1}).
\end{equation}
This follows from the fact that $f(\boldsymbol{U}^{l+1},\boldsymbol{W}^{l+1},\boldsymbol{V})$ is $L$-smooth with $L\leq L_v$, as stated in Lemma~\ref{lem:L_smooth_V}. Thus, we have
\begin{equation}\label{eq:inequality_monotonicity_3}
f(\boldsymbol{Z}^{l}_{w})\geq f(\boldsymbol{Z}^l_v).
\end{equation}

Combining~(\ref{eq:inequality_monotonicity_1}),~(\ref{eq:inequality_monotonicity_2}), and~(\ref{eq:inequality_monotonicity_3}) yields
\begin{equation}
    f(\boldsymbol{Z}^{l-1}_v)\geq f(\boldsymbol{Z}^l_u) \geq f(\boldsymbol{Z}^l_{w}) \geq f(\boldsymbol{Z}^l_v) \geq  \; \forall l \, ,
\end{equation}
which concludes the proof of monotonicity.
\end{proof}
%%%%%%%%%%%%%%%%%%%%%%%%%%%%%%%%%%%%%%%%%%%%%%%%%%%%%%%%%%%
%%%%%%%%%%%%%%%%%%%%%%%%%%%%%%%%%%%%%%%%%%%%%%%%%%%%%%%
\subsection{Existence of and convergence to a stationary point}\label{sec:convergence}
In this section, we establish that the sequence $\{\boldsymbol{U},\boldsymbol{W},\boldsymbol{V}\}^l$ generated by the matrix-inverse-free WMMSE algorithm with sufficiently small $\gamma_u > 0$ and $\gamma_v > 0$ convergences to a stationary point of~(\ref{eq:WSR_maximization}).
We state all the required lemmas, but again we relegate their proofs to the Appendix, and begin by establishing the existence of at least one limit point of $\{\boldsymbol{U},\boldsymbol{W},\boldsymbol{V}\}^l$ in~Lemma~\ref{lem:limit_point}.
\begin{lemma}\label{lem:limit_point}
The sequence of iterates generated by the matrix-inverse-free WMMSE algorithm %with \mbox{$0 <\gamma_u \leq \min(\frac{1}{L_u},\nu_u)$} and with \mbox{$ 0 < \gamma_v \leq \min(\frac{1}{L_v},\nu_v)$}, where \mbox{$L_u = \frac{2\bar{\alpha}\delta(P\kappa + \sigma^2)^2}{\sigma^2}$}, \mbox{$L_v=\frac{2\bar{\alpha}K\delta(P\kappa +d\sigma^2 )}{P\sigma^2}$}, and where $\nu_u$ and $\nu_v$ are specified in~(\ref{eq:def_nu_u}) and in~(\ref{eq:def_nu_v}), respectively,
has at least one limit point under the conditions stipulated in Theorem \ref{theo:monotonicity}.
\end{lemma}
The matrix-inverse-free WMMSE algorithm works on problem~(\ref{eq:equivalent_reformulation_tilde}) as proposed by Hu \textit{et al.}~\cite{Hu2021}. However, we are more interested in finding a stationary point of~(\ref{eq:equivalent_reformulation}) because we wish to leverage the equivalence between~(\ref{eq:equivalent_reformulation}) and the WSR maximization problem in~(\ref{eq:WSR_maximization}) established by Shi \textit{et al.}~\cite{Shi2011}. Thus, in Lemma~\ref{lem:stationary_points_equivalence}, we establish that a stationary point of~(\ref{eq:equivalent_reformulation_tilde}) is also a stationary point of~(\ref{eq:equivalent_reformulation}), after the reciprocal scaling of $\boldsymbol{U}$ and $\boldsymbol{V}$. We hasten to add that the statement of Lemma~\ref{lem:stationary_points_equivalence} was essentially given in~\cite{Hu2021}, although not formally proven.

\begin{lemma}\label{lem:stationary_points_equivalence}
If $(\overline{\boldsymbol{U}},\overline{\boldsymbol{W}},\overline{\boldsymbol{V}})$ is a stationary point of problem~(\ref{eq:equivalent_reformulation_tilde}), then $(\overline{\boldsymbol{U}}\frac{1}{\beta},\overline{\boldsymbol{W}},\overline{\boldsymbol{V}}\beta)$ is a stationary point of~(\ref{eq:equivalent_reformulation}), where \mbox{$\beta = \sqrt{\frac{P}{\sum_{m}{\mathtt{Tr}(\boldsymbol{V}_m \boldsymbol{V}^H_m)}}}$}.
\end{lemma}
We can now provide the proof of Theorem~\ref{theo:stationary_point}.
\begin{proof}
Let us denote the limit point of $\{\boldsymbol{U}\}^l$ as $\overline{\boldsymbol{U}}$, the limit point of $\{{\boldsymbol{W}}\}^l$ as $\overline{\boldsymbol{W}}$, and the limit point of $\{\boldsymbol{V}\}^l$ as $\overline{\boldsymbol{V}}$.

Let us assume to be at the limit point $\overline{\boldsymbol{X}} =(\overline{\boldsymbol{U}},\overline{\boldsymbol{W}},\overline{\boldsymbol{V}})$.
In the following, we consider a single GD step in $\boldsymbol{U}$ and $\boldsymbol{V}$ and a single Schulz iteration on $\boldsymbol{W}$, but the derivations can be easily extended to an arbitrary finite number of GD steps on $\boldsymbol{U}$ and $\boldsymbol{V}$ and of Schulz iterations on $\boldsymbol{W}$.

Let ${\boldsymbol{U}^+}$ be the next GD iterate, starting from $\overline{\boldsymbol{X}}$. By using Lemma~\ref{lem:L_smooth_U} and by considering that $f$ is convex in $\boldsymbol{U}$, we have from~\cite[Section 1.2.3]{Nesterov2014} that for any $ 0 <\gamma_u \leq \frac{1}{L_u}$
\begin{equation}
f({\boldsymbol{U}^+},\overline{\boldsymbol{W}},\overline{\boldsymbol{V}})  \leq f(\overline{\boldsymbol{U}},\overline{\boldsymbol{W}},\overline{\boldsymbol{V}}) -\frac{\gamma_u \left\lVert \nabla_u f (\overline{\boldsymbol{X}}) \right\rVert^2}{2}\raisepunct{.} 
\end{equation}
Therefore, it must hold that 
\begin{equation}\label{eq:optimality_cond_u}
\nabla_{u} f (\overline{\boldsymbol{X}}) = \boldsymbol{0}\;\; \text{and} \; \; \langle \nabla_u f(\overline{\boldsymbol{X}}),\boldsymbol{U} - \overline{\boldsymbol{U}} \rangle = 0
\end{equation}
or otherwise we would have that
\begin{equation}
    f({\boldsymbol{U}^+},\overline{\boldsymbol{W}},\overline{\boldsymbol{V}})<f(\overline{\boldsymbol{U}},\overline{\boldsymbol{W}},\overline{\boldsymbol{V}}),
\end{equation}
which would be in contradiction with the assumption that $\overline{\boldsymbol{U}}$ is a limit point of $\{ \boldsymbol{U}\}^l$. The argument for ${\boldsymbol{V}}$ is equivalent with $\gamma_v$ and $L_v$ in place of $\gamma_u$ and $L_u$, and establishes that
\begin{equation}\label{eq:optimality_cond_v}
\langle \nabla_v f(\overline{\boldsymbol{X}}),\boldsymbol{V} - \overline{\boldsymbol{V}} \rangle = 0 \, .
\end{equation}

%%%%%%%%%%%%%%%%%%%%%%%%%%%%%%%%%%%%%%%%%%%%%%%%%%%%%%%%%%%

At the limit point, the fixed point in $\boldsymbol{W}$ of the Schulz iteration must be a positive definite matrix as a singular matrix would yield $f \rightarrow \infty$ and this would contradict the monotonicity property in Theorem~\ref{theo:monotonicity}.
The only fixed point of the Schulz iteration for positive definite matrices is $\overline{\boldsymbol{W}} = \overline{\boldsymbol{E}}^{-1}$, where $\overline{\boldsymbol{E}} = \boldsymbol{E}(\overline{\boldsymbol{U}},\overline{\boldsymbol{V}})$.
Let $\mathcal{W} \triangleq \{\boldsymbol{W} \, | \, \boldsymbol{W} \succeq \boldsymbol{0}\}$ be the set of positive semidefinite matrices. Given that $\overline{\boldsymbol{E}}^{-1}$ is the optimal solution to the following convex optimization problem
\begin{subequations}
\begin{alignat}{2}
\min_{\boldsymbol{\xi}} \quad & f(\overline{\boldsymbol{U}},\boldsymbol{\xi},\overline{\boldsymbol{V}})\\
\textrm{s.t.} \quad  & \boldsymbol{\xi} \in \mathcal{W},
\end{alignat}
\end{subequations}
%where $\mathcal{W}$ is the set of positive semidefinite matrices,
it follows that 
\begin{equation}\label{eq:optimality_cond_w}
     \langle\nabla_{w} f(\overline{\boldsymbol{X}}),\boldsymbol{W} - \overline{\boldsymbol{W}}\rangle \; \geq 0 \; \; \forall\, \boldsymbol{W} \in \mathcal{W}.
\end{equation}

Combining inequalities~(\ref{eq:optimality_cond_u}),~(\ref{eq:optimality_cond_v}), and~(\ref{eq:optimality_cond_w}), we can conclude that $\overline{\boldsymbol{X}}$ is a stationary point of~(\ref{eq:equivalent_reformulation_tilde}), i.e.,
\begin{equation}
     \langle \nabla f(\overline{\boldsymbol{X}}), \boldsymbol{X} - \overline{\boldsymbol{X}}\rangle \; \geq 0 \; \; \forall\, \boldsymbol{X} \in \mathcal{W},
\end{equation}
From Lemma~\ref{lem:stationary_points_equivalence} this \mbox{implies that $\overline{\boldsymbol{X}}$ is a stationary point of~(\ref{eq:equivalent_reformulation}).}

Finally, we note that the relation between problems~(\ref{eq:WSR_maximization}) and~(\ref{eq:equivalent_reformulation}) established by Shi \textit{et al.} in~\cite[Theorem 3]{Shi2011} remains unchanged as it not determined by the type of algorithm used to address~(\ref{eq:equivalent_reformulation}). Therefore, in order to show that $\overline{\boldsymbol{V}}$ is a stationary point of~(\ref{eq:WSR_maximization}) if and only if $\overline{\boldsymbol{X}}=(\overline{\boldsymbol{U}},\overline{\boldsymbol{W}},\overline{\boldsymbol{V}})$ is a stationary point of~(\ref{eq:equivalent_reformulation}) for some $\overline{\boldsymbol{U}}$ and $\overline{\boldsymbol{W}}$, we can verbatim follow the second part of the proof in~\cite[Theorem 3]{Shi2011}.
\end{proof}
%%%%%%%%%%%%%%%%%%%%%%%%%%%%%%%%%%%%%%%%%%%%%%%%%%

\section{Deep-unfolding-based implementation}
In order to exemplify the use of the matrix-inverse-free WMMSE algorithm and boost its performance within a fixed number of algorithm iterations, we consider a deep-unfolding-based implementation.
By replicating the structure of an iterative algorithm into a network architecture and by embedding trainable modules in it, deep unfolding enables optimization of the performance within a fixed number of iterations~\cite{Monga2021}.
\subsection{Network architecture}
To fully exploit the potential of deep unfolding, we propose a more flexible variant of the matrix-inverse-free WMMSE algorithm. In particular, we have empirically observed that the bounds on $\gamma_u$ and $\gamma_v$ given in Theorem~\ref{theo:monotonicity} are overly conservative in practice and the matrix-inverse-free WMMSE algorithm works well for much more liberally chosen step sizes. We thus propose to choose more aggressive step sizes.
Specifically, we explicitly compute the optimal step sizes $\gamma^*_u$ and $\gamma^*_v$, which give the largest decrement in the cost function, and take steps of length $\gamma_u = \phi_u \gamma^*_u$ and $\gamma_v = \phi_v \gamma^*_v$, where $\phi_u \in (0,2)$ and $\phi_v \in (0,2)$ become trainable parameters. In general, computing $\gamma^*_u$ and $\gamma^*_v$ is not trivial, but we can exploit the quadratic nature of $f$ as a function of $\gamma_u$ ($\gamma_v$) and straightforwardly compute a closed-form expression of $\gamma^*_{u}$ ($\gamma^*_{v}$). In particular, any $\gamma_u$ ($\gamma_v$) in the range $(0,2\gamma^*_u)$ ($(0,2\gamma^*_v)$), will guarantee a decrement of $f$.
%We introduce as trainable parameter a pre-multiplier $\phi_u$ ($\phi_v$) of the optimal step size $\gamma^*_u$ ($\gamma^*_v$). We restrict it to be in the range $[0,2]$ to ensure that the overall step size given by $\phi_u\gamma^*_u$ ($\phi_v\gamma^*_v$) is in the range $(0,2\gamma^*_u]$ ($(0,2\gamma^*_v]$), while granting more
This provides more flexibility to the network and makes the network invariant to the absolute scaling of the channel and the noise power.
However, by adopting more aggressive step sizes we might occasionally violate not only~(\ref{eq:condition_eig_EW_delta}), which guarantees monotonicity of the sequence of iterates over the updates of $\boldsymbol{W}$, but also $\lambda_{\mathrm{max}}(\boldsymbol{E}_k\boldsymbol{W}_k) \leq 2$ $\forall k$. If the latter condition does not hold, then Schulz iterations might yield indefinite $\boldsymbol{W}$ and this would break the matrix-inverse-free WMMSE algorithm. Thus, we propose to re-normalize $\boldsymbol{W}$ in each algorithm iteration before updating $\boldsymbol{W}$, i.e, 
\begin{equation}\label{eq:Schulz_spectral_radius}
    \boldsymbol{W}_k = \frac{\boldsymbol{W}^{\mathrm{end}}_k}{\eta \big({\boldsymbol{E}}_k \boldsymbol{W}^{\mathrm{end}}_k\big)} \; \mathrm{for} \; k = 1,\cdots,K,
\end{equation}
where $\boldsymbol{W}^{\mathrm{end}}_k$ is the output of the last Schulz iteration in the previous algorithm iteration and $\eta(\boldsymbol{X})$ is a tight upper bound on the spectral radius of square matrix $\boldsymbol{X}$ of dimension $m$, defined as $\eta(\boldsymbol{X}) \triangleq \max_{i} \sqrt{b_i}$, where $b_i = \sum_{j=1}^{m} |\boldsymbol{X}_{i,j}| c_j$ for $i=1,\cdots,m$, and $c_j = \sum_{i=1}^{m} | \boldsymbol{X}_{i,j} |$ for $j=1,\cdots,m$~\cite{Byrne2009}.
This re-normalization may violate monotonicity but ensures the correct operation of the algorithm. Namely, it ensures that Schulz iterations always yield PSD $\boldsymbol{W}$. Moreover, the re-normalization ensures that if $\boldsymbol{W}$ is already very close to the inverse of $\boldsymbol{E}$ so that $\boldsymbol{EW} \approx \boldsymbol{I}$, then $\eta(\boldsymbol{EW}) \approx 1$ and the perturbation of $\boldsymbol{W}$ is minimal. 

Further, to confer even more flexibility to the network, we incorporate Nesterov acceleration~\cite{Nesterov1938}, a well-known technique to speed up convergence of first-order methods, into the update steps of $\boldsymbol{U}$~(\ref{eq:GD_u}) and $\boldsymbol{V}$~(\ref{eq:GD_v}). Nesterov acceleration introduces a momentum parameter $\theta$, which 
we treat as trainable. In addition, as in~\cite{Pellaco2021}, we introduce an extra learnable parameter $\xi$, which acts on the gradient input. This leads to the following trainable updates of $\boldsymbol{U}$ and $\boldsymbol{V}$, respectively, %~\cite{Nakerst2020}
%The trainable gradient updates for $\boldsymbol{U}$ and $\boldsymbol{V}$ are, respectively,
\begin{equation}\label{eq:GD_u_trainable}
\boldsymbol{U}_k^{(j)} = {\boldsymbol{U}_k}^{(j-1)} + \theta_u{{\bar{\boldsymbol{U}}}_k}^{(j-1)} - \phi_u{\gamma^*_u}\nabla_u f\big({{\boldsymbol{U}_k}}^{(j-1)}+\xi_u{{\bar{\boldsymbol{U}}}}_k^{(j-1)}\big)
\end{equation}
and
\begin{equation}\label{eq:GD_v_trainable}
\boldsymbol{V}_k^{(j)} = {\boldsymbol{V}_k}^{(j-1)} + \theta_v{{\bar{\boldsymbol{V}}_k}}^{(j-1)} - \phi_v{\gamma^*_v}\nabla_v f\big({{\boldsymbol{V}_k}}^{(j-1)}+\xi_v{{\bar{\boldsymbol{V}}}_k}^{(j-1)}\big),
\end{equation}
for $k = 1,\cdots,K$, where the superscript $j$ indicates the variable given by the $j^{th}$ GD step, where $\nabla_u f$ and $\nabla_v f$ are defined in~(\ref{eq:gradient_u_expression}) and~(\ref{eq:gradient_v_expression}), respectively, and where $\bar{\boldsymbol{U}}_k^{(j)}$ and $\bar{\boldsymbol{V}}_k^{(j)}$ are defined as $\bar{\boldsymbol{U}}_k^{(j)} \triangleq \boldsymbol{U}_k^{(j)} - \boldsymbol{U}_k^{(j-1)}$ and $\bar{\boldsymbol{V}}_k^{(j)} \triangleq \boldsymbol{V}_k^{(j)} - \boldsymbol{V}_k^{(j-1)}$, respectively. 
We collectively indicate the pre-multiplier parameters across the GD steps on $\boldsymbol{U}$ and $\boldsymbol{V}$ and across the algorithm iterations as $\boldsymbol{\Phi} = [\boldsymbol{\Phi}_u,\boldsymbol{\Phi}_v]$ and we adopt analogous notation for momentum parameters $\theta$ and $\xi$, i.e., $\boldsymbol{\Theta} = [\boldsymbol{\Theta}_u,\boldsymbol{\Theta}_v]$ and $\boldsymbol{\Xi} = [\boldsymbol{\Xi}_u, \boldsymbol{\Xi}_v]$, respectively.

Finally, we have observed experimentally that it suffices to scale $\boldsymbol{V}$ only after $J_v$ steps (i.e. not after each single GD step) and that the reciprocal scaling of $\boldsymbol{U}$ can be omitted. Although this might also break monotonicity of the sequence of iterates, we adopt these simplifications to further reduce the complexity of the implemented algorithm. 
Furthermore, we execute $J_w$ Schulz iterations as part of the initialization. This is done to warm start the algorithm because a trivial initialization of $\boldsymbol{W}$, like scaled identity, would severely penalize the first update of $\boldsymbol{U}$ (in which $\nabla_u f$ depends on $\boldsymbol{W}$) and would in turn penalize the WSR achieved in the first algorithm iteration. This practice of warm starting to reach efficiency is common in optimization, but we stress that we do not utilize inverses in the warm start as instead done in \cite{Hu2021}.
Algorithm~\ref{alg:WMMSE_for_unfolding} summarizes this more flexible variant of the
matrix-inverse-free WMMSE algorithm formulated to leverage deep unfolding. We build a network that replicates a finite number of iterations of Algorithm~\ref{alg:WMMSE_for_unfolding} and we refer to it as \textit{unfolded matrix-inverse-free WMMSE network}.
It must be mentioned that although Theorem~\ref{theo:monotonicity} and Theorem~\ref{theo:stationary_point} do not apply to this case, experimental results (see Section~\ref{sec:numerical_results}) show that the unfolded matrix-inverse-free WMMSE network reaches satisfactory results and shows a practically monotonic behavior. 
\begin{algorithm}
\caption{Matrix-inverse-free WMMSE for unfolding}\label{alg:WMMSE_for_unfolding}
\begin{algorithmic}
\State {Initialize $\boldsymbol{V}$ such that $\sum_{m=1}^{K}{\mathtt{Tr}(\boldsymbol{V}_m\boldsymbol{V}_m^H) = P}$}
\State {Initialize $\boldsymbol{U}$}
\State {Initialize $\boldsymbol{W} = {\boldsymbol{I}}/{\mathtt{Tr}(\boldsymbol{E})}$}
\For {$j=1,\cdots,J_w$}
\State{$\boldsymbol{W}_k \gets \boldsymbol{W}_k(2\boldsymbol{I} - {\boldsymbol{E}}_k\boldsymbol{W}_k) \; \mathrm{for} \; k =1,\cdots,K$}
\EndFor
\For{$l=1,\cdots,L$}
\For {$j=1,\cdots,J_u$}
\State{Update $\boldsymbol{U}$~(\ref{eq:GD_u_trainable}) }
\EndFor
\State{$\boldsymbol{W}_k \gets {\boldsymbol{W}_k}/{\eta \big({\boldsymbol{E}}_k \boldsymbol{W}_k\big)} \; \mathrm{for} \; k = 1,\cdots,K$}
\For {$j=1,\cdots,J_w$}
\State{$\boldsymbol{W}_k \gets \boldsymbol{W}_k(2\boldsymbol{I} - {\boldsymbol{E}}_k\boldsymbol{W}_k) \; \mathrm{for} \; k =1,\cdots,K$}
\EndFor
\For {$j=1,\cdots,J_v$}
\State{Update $\boldsymbol{V}$~(\ref{eq:GD_v_trainable}) }
\EndFor
\State $\boldsymbol{V}_k \gets \boldsymbol{V}_k\sqrt{\frac{P}{\sum_{m}{\mathtt{Tr}(\boldsymbol{V}_m \boldsymbol{V}^H_m)}}}$ for $k = 1, \cdots, K$
\EndFor
\end{algorithmic}
\end{algorithm}

\subsection{Training}
We ultimately do not aim to approximate the update equations of the WMMSE algorithm nor achieve the same WSR as the WMMSE algorithm (which is suboptimal in general).
Our goal is to attain the highest possible WSR.
Therefore, as in~\cite{Pellaco2021}, we adopt a loss function which directly depends on the WSR achieved by the beamformer given as output by the network. However, in order to prevent training difficulties, we include also the WSR achieved by the beamformers
given as output by all the other layers. This idea follows from the auxiliary classifiers introduced in~\cite{Szegedy2015} and was recommended by~\cite{Wiesel2019}. Thus, we adopt this loss function
\begin{equation}\label{eq:loss_function}
\mathcal{L}(\boldsymbol{\Phi,\Theta,\Xi}) = -\frac{1}{N_{\mathrm{s}}} \sum_{n= 1}^{N_{\mathrm{s}}}{\sum_{l = 1}^{L}{f_{\mathrm{WSR}}\big(\boldsymbol{H}^{n}},\boldsymbol{V}^{l,n}}(\boldsymbol{\Phi},\boldsymbol{\Theta}, \boldsymbol{\Xi})\big),   
\end{equation}
where $N_{\mathrm{s}}$ is the number of channel realizations in the training set and $f_{\mathrm{WSR}}(\boldsymbol{H}^{n},\boldsymbol{V}^{l,n}\big(\boldsymbol{H}^{n};\boldsymbol{\Phi},\boldsymbol{\Theta}, \boldsymbol{\Xi})\big)$ is the WSR~(\ref{eq:WSR_maximization_cost_function}) attained  with $\boldsymbol{V}^{l,n}$, the beamformer given as output by the $l^{th}$ layer of the network (with trainable parameters $\boldsymbol{\Phi},\boldsymbol{\Theta}, \boldsymbol{\Xi}$) when the $n^{th}$ channel realization $\boldsymbol{H}^{n}$ is given as input.
In addition to the benefit of unsupervised training, loss function~(\ref{eq:loss_function}) allows us to actually outperform the WMMSE algorithm for a finite number of iterations (see Section~\ref{sec:numerical_results}). We adopt the Adam optimizer~\cite{Adam2014} to minimize~(\ref{eq:loss_function}).
Finally, the WSR computation in the loss function does involve a matrix inverse, see~(\ref{eq:R}). However, the loss function is only used in the training phase, and hence no matrix inverse operations are required in deployment phase.

\section{Numerical results}\label{sec:numerical_results}
\subsection{Setup}
We implemented the WMMSE algorithm in MATLAB and the unfolded matrix-inverse-free WMMSE network in Python 3.6.8 with Tensorflow 1.13.1. The code to reproduce the results is available at~\cite{GITHUB}.

We used the following settings for the numerical simulations:
\begin{itemize}
    \item We set $\alpha_k = 1$ for $i=1,\cdots,K$ and $\frac{P}{\sigma^2}=10$ dB.

    \item In the WMMSE algorithm, we stop the bisection search when the constraint on the transmit power is met with equality with an error less than or equal to $10^{-4}$ and we assume that convergence is reached when the increment in WSR is less than or equal to $10^{-4}$ bits per channel use at the next algorithm iteration.   
    
    \item We initialize the step size pre-multipliers $\phi_u$ and $\phi_v$ to one and the momentum parameters $\theta_u$, $\xi_u$ and $\theta_v$, $\xi_v$ to zero across all GD steps and algorithm iterations. 
    
    \item In the WMMSE algorithm and in the unfolded matrix-inverse-free WMMSE network, we initialize $\boldsymbol{V}_k$ as $\boldsymbol{V}_k = \widetilde{\boldsymbol{H}}^H_k$ for $k=1,\cdots,K$, where $\widetilde{\boldsymbol{H}}_k $ is the matrix given by the $d$ rows of $\boldsymbol{H}_k$ with largest power. Then, we normalize $\boldsymbol{V}$ such that the power constraint is met with equality.
    
    \item In the unfolded matrix-inverse-free WMMSE network, after the initialization of $\boldsymbol{V}$, we initialize $\boldsymbol{U}_k$ as $\boldsymbol{U}_k = \rho_k^*\boldsymbol{I}$ for $k=1,\cdots,K$, where $\rho_k^* =\argmax_{\rho_k}{\mathtt{Tr}\big({\boldsymbol{E}}_k(\rho_k\boldsymbol{I},\boldsymbol{V})\big)}$.

    \item In the unfolded matrix-inverse-free WMMSE network, in the first GD step, we set $ \boldsymbol{V}_k^{l,-1} = \boldsymbol{0}$ and $\boldsymbol{U}_k^{l,-1} = \boldsymbol{0}$ for $k=1,\ldots,K$ and for $l=1,\ldots,L$.    
    
    \item We fix the learning rate of the Adam optimizer to $10^{-3}$. 
    
    \item We apply the following complex-to-real  mapping because current software tools do not support complex variable optimization, i.e., $\boldsymbol{x}^{\prime} = [\Re(\boldsymbol{x})^T, \ \Im(\boldsymbol{x})^T]^T$,
    where $\boldsymbol{x}$ is a complex column vector and $$
    \boldsymbol{X}^{\prime} = \begin{bmatrix} \Re(\boldsymbol{X}) & -\Im(\boldsymbol{X}) \\ \Im(\boldsymbol{X}) & \Re(\boldsymbol{X}) \end{bmatrix},
    $$ where $\boldsymbol{X}$ is a complex matrix, and $\Re(\cdot)$ and $\Im(\cdot)$ denote the real and the imaginary parts.
    
    \item We can easily impose the range constraint on step size pre-multipliers $\phi_u$ and $\phi_v$ across all GD steps and algorithm iterations by implementing them as scaled outputs of sigmoid functions.
    
    \item We draw i.i.d.\ complex-valued (rotationally symmetric) Gaussian (Rayleigh) channel gains with zero mean and unit variance.
    %from $\mathcal{CN}(\boldsymbol{0},\,\boldsymbol{I})$.
    The training and the test sets consist of $10^6$ and $10^5$ channel realizations, respectively, and we fix the batch size to $10^2$. 
    
    \item We train a different network for each combination of $K$, $N$, and $L$.

    \item For IAIDNN~\cite{Hu2021}, we use the authors' code available at~\cite{GITHUB_IAIDNN}. In this case, we initialize $\boldsymbol{V}$ with zero forcing as proposed by the authors, we fix the batch size to $10^2$, and we train the network until convergence.

\end{itemize}

\subsection{Results}
In this section, we compare i) the performance of the unfolded matrix-inverse-free WMMSE network, ii) the performance of the WMMSE algorithm truncated to the same number of iterations, and iii) the performance of the IAIDNN algorithm proposed in~\cite{Hu2021}.
% We should make the distinction between fully loaded and lightly loaded scenarios
We consider two fully loaded scenarios, i.e. $(M=8, K=2, N=4, d=4)$ and $(M=8, K=4, N=2, d=2)$, and one lightly loaded scenario, i.e., $(M=8, K=2, N=2, d=2)$.
%%%%%%%%%%%%%%%%%%%%%%%%%%%%%%%%%%%%%%%%%%%%%%%%%%
Clearly, the performance of the unfolded matrix-inverse-free WMMSE network depends on the selected number of GD steps and of Schulz iterations. Although a thorough investigation is outside the scope of the paper, we have carried out preliminary tests and derived i) that for the lightly loaded scenario and for the fully loaded scenario with $d=2$ it is more beneficial to take more GD steps on $\boldsymbol{V}$ rather than on $\boldsymbol{U}$, ii) that for the fully loaded scenario with $d=4$ it is more beneficial to take more GD steps on $\boldsymbol{U}$ rather than on $\boldsymbol{V}$ and iii) that for the $\boldsymbol{W}$ update it is convenient to only perform 2 Schulz iterations rather than more in all scenarios. Thus, to evaluate the performance of the unfolded matrix-inverse-free WMMSE network we select different triplets $(J_u,J_w,J_v)$ with increasing computational complexity.

\begin{figure}[ht]
         \centering
         \scalebox{0.85}{\includegraphics{{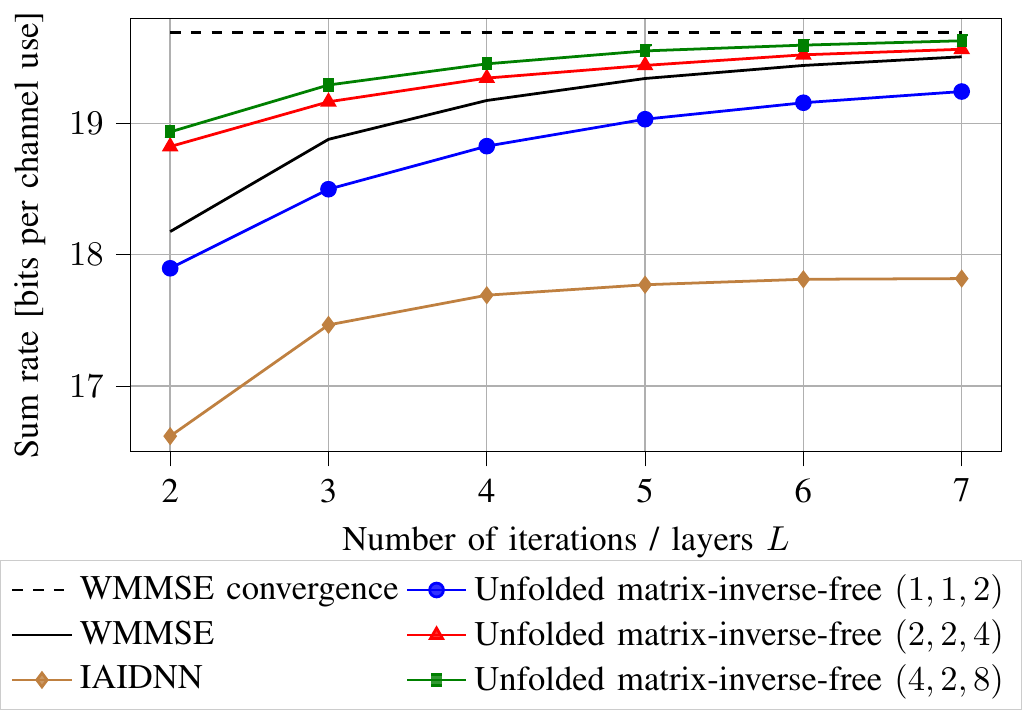}}}
		\caption{WSR obtained with $M = 8,K=4, N=2, d=2$.}
         \label{fig:M8_K4_N2_d2}   
\end{figure}

\begin{figure}[ht]
         \centering
         \scalebox{0.85}{\includegraphics{{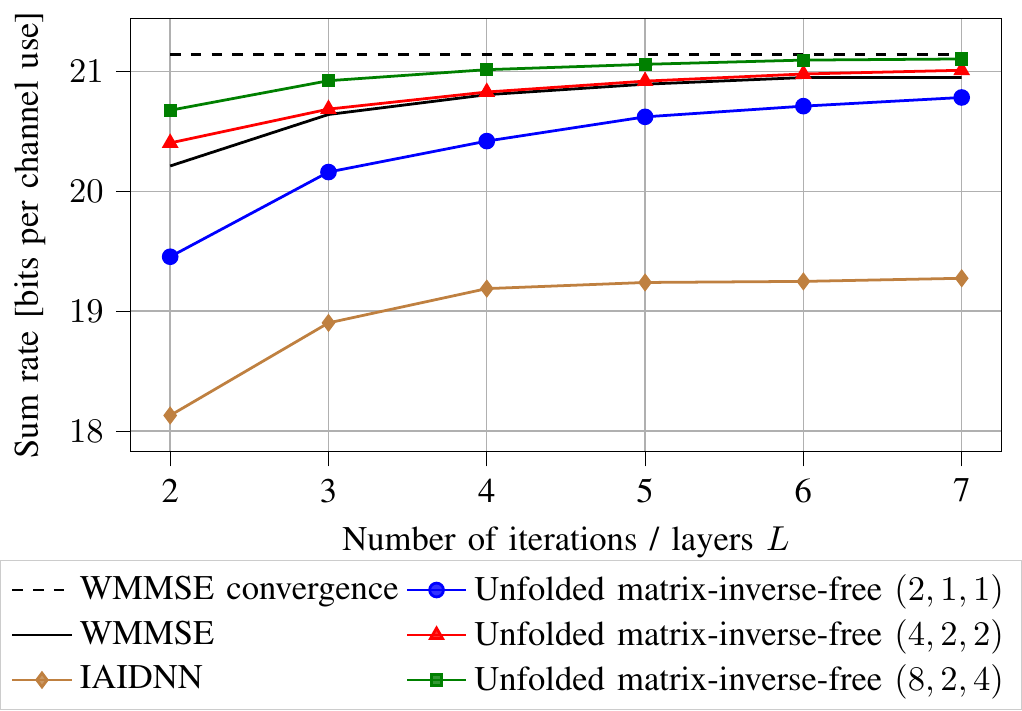}}}
		\caption{WSR obtained with $M = 8,K=2, N=4, d=4$. }
         \label{fig:M8_K2_N4_d4}   
\end{figure}

\begin{figure}[ht]
         \centering
         \scalebox{0.85}{\includegraphics{{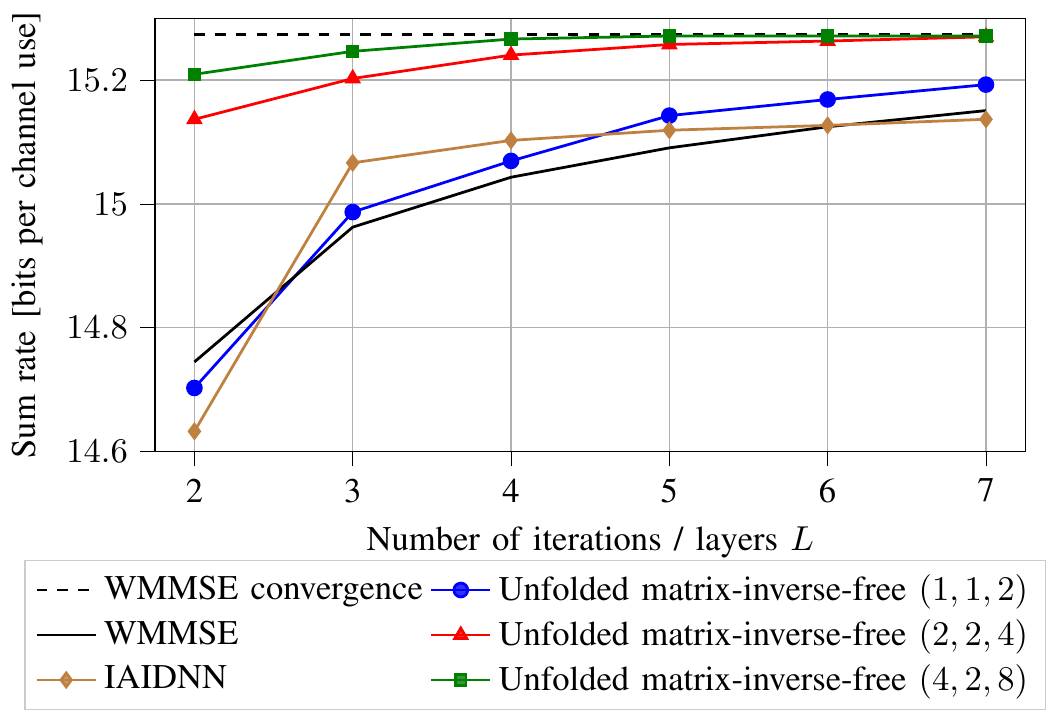}}}
		\caption{WSR obtained with $M = 8,K=2, N=2, d=2$.}
         \label{fig:M8_K2_N2_d2}   
\end{figure}
%%%%%%%%%%%%%%% Figures%%%%%%%%%%%%%%%%%%%
Figures~\ref{fig:M8_K4_N2_d2} and~\ref{fig:M8_K2_N4_d4} show the performance in the two fully loaded scenarios. As $J_u$, $J_w$, and $J_v$ increase the attained WSR increases as well, as expected. The unfolded matrix-inverse-free WMMSE network with triplets $(1,1,2)$ and $(2,1,1)$ is surpassed by the WMMSE algorithm truncated to the same number of iterations, whereas with the other triplets the unfolded matrix-inverse-free WMMSE network outperforms the truncated WMMSE algorithm. With triplets $(4,2,8)$ and $(8,2,4)$, it even achieves 99 percent of the WSR achieved by the WMMSE algorithm at convergence.
On the contrary, the IAIDNN reaches 91 percent of the WSR at convergence. This was expected as the authors of~\cite{Hu2021} recognize that the IAIDNN struggles in case of fully loaded scenarios because of the adopted Taylor-expansion-based approximation of the matrix inverse.
Figure~\ref{fig:M8_K2_N2_d2} shows the performance of the various approaches in the lightly loaded scenario (note the narrow range of WSR). The same conclusions of the fully loaded scenarios hold, but with a significant difference. In this case, the unfolded matrix-inverse-free WMMSE network with triplet $(1,1,2)$ surpasses the truncated WMMSE algorithm and the IAIDNN yields an excellent performance as it reaches 99 percent of the WMMSE algorithm at convergence for $L=7$.
%%%%%%%%%%%%%%%%%%%%%%%%%%%%%%%%%%%%%%%%%%%%%%%%%%%%%%%%%%%%%%%%%

\section{Conclusion}
We proposed the first matrix-inverse-free formulation of the WMMSE algorithm for the MU-MIMO case with i) provable convergence to a stationary point of the NP-hard WSR maximization problem and ii) compliance with real-time implementation requirements. By unfolding and training a finite number of learnable iterations of the matrix-inverse-free WMMSE algorithm, we showed that replacing matrix inverses with trainable accelerated gradient descent steps and with Schulz iterations does not degrade the performance, yet accelerates convergence. 
Although we considered a single base station, by appropriately changing the power constraint, our method can be applied mutatis mutandis to the multiple base station case.
We focused on the popular WMMSE, but our approach is in principle applicable to a large variety of algorithms that involve matrix inverses, eigendecompositions, and in general non parallelizable operations that make such algorithms not relevant for time-critical applications.

{\appendix
%%%%%%%%%%%%%%%%%%%%%%%%%%%%%%%%%%%%%%%%%%%%%%%%%%%%%%%%%%%%%%%%%%%%%%%
From~(\ref{eq:condition_eig_EW_delta}) and the structure of $\boldsymbol{E}_k$~(\ref{eq:E_tilde}), it follows that $\mathtt{Tr}(\boldsymbol{W}_k \boldsymbol{U}^H_k \boldsymbol{U}_k) \leq \frac{\mathtt{Tr}(\boldsymbol{W}_k \boldsymbol{E}_k)}{\sigma^2} \leq \frac{ d \delta}{\sigma^2}$ $\forall k$ and that
$\lambda_{\mathrm{max}}(\boldsymbol{W}_k \boldsymbol{U}^H_k \boldsymbol{U}_k) \leq \frac{\lambda_{\mathrm{max}}(\boldsymbol{W}_k \boldsymbol{E}_k)}{\sigma^2} \leq \frac{ \delta}{\sigma^2}$ $\forall k$,
where we have used $\sum_{m=1}^{K}{\mathtt{Tr}(\boldsymbol{V}_m \boldsymbol{V}^H_m)}=P$ and $\boldsymbol{W}_k \succeq 0 $. This implies that
\mbox{$\lVert \boldsymbol{W}_k^{\frac{1}{2}} \boldsymbol{U}_k^H\rVert_2 \leq \frac{\sqrt{\delta}}{\sigma},
$}
and that
\mbox{$ \lVert\boldsymbol{U}_k \boldsymbol{W}_k \boldsymbol{U}^H_k\rVert_2  \leq \frac{\delta}{\sigma^2} \, \forall k.$}
%%%%%%%%%%%%%%%%%%%%%%%%%%%%%%%%%%%%%%%%%%%%%%%%%%%%%%%%%%%%%%%%%%%%%%%
\subsection{Proof of Lemma~\ref{lem:lower_bound_E}}
\begin{proof}
Let $\eta = \sum_m \|\boldsymbol{V}_m\|^2_F = \sum_m \mathtt{Tr}(\boldsymbol{V}_m \boldsymbol{V}^H_m )$. Then
\begin{equation}
\|\boldsymbol{H}_k \boldsymbol{V}_k\|^2_{F} \leq \kappa \| \boldsymbol{V}^H_k \|^2_F \leq \kappa \sum_m \|\boldsymbol{V}_m\|^2_{F} = \kappa \eta.
\end{equation}

From the structure of $\boldsymbol{E}_k $~(\ref{eq:E_tilde}), it follows that $\lambda_{\mathrm{min}}(\boldsymbol{E}_k) \geq \lambda_{\mathrm{min}}(\boldsymbol{E}_k - \!\sum_{m\neq k}{\!\boldsymbol{U}^H_k \boldsymbol{H}_k \boldsymbol{V}_m  \boldsymbol{V}^H_m \boldsymbol{H}^H_k \boldsymbol{U}_k )}$. Thus, for the purpose of this proof, we consider $\boldsymbol{E}_k$ without the interference term $\sum_{m\neq k}{\!\boldsymbol{U}^H_k \boldsymbol{H}_k \boldsymbol{V}_m  \boldsymbol{V}^H_m \boldsymbol{H}^H_k \boldsymbol{U}_k }$.
Let us consider $\boldsymbol{a}^H \boldsymbol{E}_k \boldsymbol{a}$ for some $\boldsymbol{a}$ where $\|\boldsymbol{a}\|_2 = 1$ and let $\boldsymbol{b} = \boldsymbol{U}\boldsymbol{a}$. Then it follows that
\begin{equation}
\boldsymbol{a}^H \boldsymbol{E}_k \boldsymbol{a} = \| \boldsymbol{a} - \boldsymbol{V}_k^H \boldsymbol{H}_k^H \boldsymbol{b}\|_2^2 + \frac{\eta \sigma^2}{P} \| \boldsymbol{b} \|_2^2.
\end{equation}
If we assume that $\|\boldsymbol{b}\|_2^2 \geq \frac{1}{\kappa \eta}$ it follows immediately that
\begin{equation}
\boldsymbol{a}^H \boldsymbol{E}_k \boldsymbol{a} \ge \frac{\eta \sigma^2}{P} \| \boldsymbol{b} \|_2^2 \ge \frac{\sigma^2}{P \kappa} \ge \frac{\sigma^2}{P \kappa + \sigma^2}\raisepunct{.}
\end{equation}
If we instead assume that $\|\boldsymbol{b}\|_2^2 \le \frac{1}{\kappa \eta}$ or equivalently that $\|\boldsymbol{b}\|_2 \le \frac{1}{\sqrt{\kappa \eta}}$ it follows that
\begin{subequations}
\begin{alignat}{2}
\| \boldsymbol{a} - \boldsymbol{V}_k^H \boldsymbol{H}_k^H \boldsymbol{b}\|_2 &\geq \| \boldsymbol{a} \|_2 - \| \boldsymbol{V}_k^H \boldsymbol{H}_k^H \boldsymbol{b} \|_2 \\ &\geq \| \boldsymbol{a} \|_2 - \sqrt{\kappa \eta} \| \boldsymbol{b} \|_2 =
1 - \sqrt{\kappa \eta} \| \boldsymbol{b} \|_2 \geq 0 ,
\end{alignat}
\end{subequations}
which implies that
\begin{equation}
\| \boldsymbol{a} - \boldsymbol{V}_k^H \boldsymbol{H}_k^h \boldsymbol{b}\|_2^2 \ge \big( 1 - \sqrt{\kappa \eta} \| \boldsymbol{b} \| _2\big)^2 \!= \!1 - 2\sqrt{\eta \kappa} \|\boldsymbol{b}\|_2 + \eta \kappa \|\boldsymbol{b}\|_2^2
\end{equation}
and
\begin{subequations}
\begin{alignat}{2}
\boldsymbol{a}^H \boldsymbol{E}_k \boldsymbol{a} &= \| \boldsymbol{a} - \boldsymbol{V}_k^H \boldsymbol{H}_k^h \boldsymbol{b}\|_2^2 + \frac{\eta \sigma^2}{P} \|\boldsymbol{b}\|_2^2 \\ &\geq 1 - 2\sqrt{\eta \kappa} \|\boldsymbol{b}\|_2 + \eta \kappa \|\boldsymbol{b}\|_2^2 + \frac{\eta \sigma^2}{P} \|\boldsymbol{b}\|_2^2.
\end{alignat}
\end{subequations}
Minimizing the right hand side over the scalar $\|\boldsymbol{b}\|_2$ yields
\begin{equation}
\boldsymbol{a}^H \boldsymbol{E}_k \boldsymbol{a} \ge \frac{\eta \sigma^2 / P}{\eta \sigma^2 / P + \kappa \eta} = \frac{\sigma^2}{ P\kappa + \sigma^2}\raisepunct{.}
\end{equation}
Thus, we have
$\boldsymbol{a}^H \boldsymbol{E}_k \boldsymbol{a} \ge \frac{\sigma^2}{P\kappa + \sigma^2}$
for any $\boldsymbol{a}$ where $\|\boldsymbol{a}\|_2 = 1$,
\begin{equation}
\sigma_{\min}(\boldsymbol{E}_k) = \min_{\|\boldsymbol{a}\| = 1} \, \boldsymbol{a}^H \boldsymbol{E}_k \boldsymbol{a} \ge \frac{\sigma^2}{P\kappa + \sigma^2}\raisepunct{,}
\end{equation}
and consequently $\lambda_{\mathrm{min}}(\boldsymbol{E}_k) = \sigma_{\min}(\boldsymbol{E}_k) \ge \frac{\sigma^2}{P\kappa + \sigma^2} \raisepunct{.}$
\end{proof}
%%%%%%%%%%%%%%%%%%%%%%%%%%%%%%%%%%%%%%%%%%%%%%%%%%%%%%%%
\subsection{Proof of Lemma~\ref{lem:on_eigenvalues_of_EW}}
\begin{proof}
From $\boldsymbol{W} \succeq 0$, we have
$   \lambda_i(\boldsymbol{EW}) = \lambda_i(\boldsymbol{W}^{\frac{1}{2}} \boldsymbol{E}\boldsymbol{W}^{\frac{1}{2}} )$ $\forall i$,
where we have applied the property that the eigenvalues of a product of matrices are invariant under cyclic permutation
of the product order. Given that $\boldsymbol{E} \succeq 0$, $\boldsymbol{W}^{\frac{1}{2}} \boldsymbol{E}\boldsymbol{W}^{\frac{1}{2}}$ is positive semidefinite and hence it has real-valued and non-negative eigenvalues and $\lambda_i(\boldsymbol{W}^{\frac{1}{2}} \boldsymbol{E}\boldsymbol{W}^{\frac{1}{2}}) = \sigma_i(\boldsymbol{W}^{\frac{1}{2}} \boldsymbol{E}\boldsymbol{W}^{\frac{1}{2}}) \; \forall i$. It follows that $\boldsymbol{EW}$ has real-valued and non-negative eigenvalues as well and \mbox{$\lambda_i(\boldsymbol{E} \boldsymbol{W}) = \sigma_i(\boldsymbol{W}^{\frac{1}{2}} \boldsymbol{E} \boldsymbol{W}^{\frac{1}{2}})$} $\forall i $.
\end{proof}
%%%%%%%%%%%%%%%%%%%%%%%%%%%%%%%%%%%%%%%%%%%%%%%%%%%%%%%%
\subsection{Proof of Lemma~\ref{lem:L_smooth_U}}
\begin{proof}
We define $f(\boldsymbol{U}) \triangleq f(\boldsymbol{U},\boldsymbol{W},\boldsymbol{V})$.
We have that
\begin{equation}
\lVert\nabla_u f(\dot{\boldsymbol{U}}_k) - \!\nabla_u f(\ddot{\boldsymbol{U}}_k)\rVert_F \leq \lVert\boldsymbol{G}_k\rVert_2 \lVert\dot{\boldsymbol{U}}_k - \! \ddot{\boldsymbol{U}}_k\rVert_F \lVert\boldsymbol{W}_k\rVert_2,
\end{equation}
where
\begin{equation}
\boldsymbol{G}_k = 2\alpha_k\Big( \boldsymbol{H}_k \sum_{m=1}^{K}{\!\boldsymbol{V}_m{\boldsymbol{V}_m}^H\boldsymbol{H}^H_k} + \boldsymbol{I}\frac{\sigma^2}{P}\sum_{m = 1}^{K}{\mathtt{Tr}( \boldsymbol{V}_m  {\boldsymbol{V}}^{H}_m )}\Big).
\end{equation}
This implies that
$\lVert\boldsymbol{G}_k\rVert_2 \leq 2 \bar{ \alpha} (P\kappa + \sigma^2) \,\forall k$
and leads to the following upper bound for any $\dot{\boldsymbol{U}}$ and $\ddot{\boldsymbol{U}}$ 
\begin{equation}
\lVert\nabla_u f(\dot{\boldsymbol{U}}_k) - \nabla_u f(\ddot{\boldsymbol{U}}_k)\rVert_F \leq L_u \lVert\dot{\boldsymbol{U}}_k - \ddot{\boldsymbol{U}}_k\rVert_F \; \; \forall k,
\end{equation}
where $L_u = \frac{2\bar{\alpha}\delta(P\kappa + \sigma^2)^2}{\sigma^2}\raisepunct{.}$
\end{proof}
%%%%%%%%%%%%%%%%%%%%%%%%%%%%%%%%%%%%%%%%%%%%%%%%%%%%%%%%
\subsection{Proof of Lemma~\ref{lem:L_smooth_V}}
\begin{proof}
We define $f(\boldsymbol{V}) \triangleq f(\boldsymbol{U},\boldsymbol{W},\boldsymbol{V})$.
We have that
\begin{equation}
\lVert\nabla_v f(\dot{\boldsymbol{V}}_k) - \nabla_v f(\ddot{\boldsymbol{V}}_k)\rVert_F \leq \lVert\boldsymbol{F}_k\rVert_2 \lVert\dot{\boldsymbol{V}}_k - \ddot{\boldsymbol{V}}_k\rVert_F,  
\end{equation}
where 
\begin{align}
    \boldsymbol{F}_k &= 2\sum_{m=1}^{K}{\alpha_m \boldsymbol{H}^H_m \boldsymbol{U}_m \boldsymbol{W}_m {\boldsymbol{U}}^H_m  \boldsymbol{H}_m } \\ &+2 \boldsymbol{I}  \frac{\sigma^2}{P}\sum_{m = 1}^{K}{\alpha_m \mathtt{Tr}( \boldsymbol{W}_m  {\boldsymbol{U}}^{H}_m  \boldsymbol{U}_m )}. \nonumber
\end{align}
It follows that $\lVert\boldsymbol{F}_k \rVert_2 \leq L_u\, \forall k$, where $L_v = \frac{2\bar{\alpha}K\delta(P\kappa + d\sigma^2)}{P\sigma^2}\raisepunct{.}$
Thus, we have the following upper bound for any $\dot{\boldsymbol{V}}$ and $\ddot{\boldsymbol{V}}$
\begin{equation}
\lVert\nabla_v f(\dot{\boldsymbol{V}}_k) - \nabla_v f_v(\ddot{\boldsymbol{V}}_k)\rVert_F \leq L_u \lVert\dot{\boldsymbol{V}}_k - \ddot{\boldsymbol{V}}_k\rVert_F \;\; \forall k.
\end{equation}
\end{proof}
%%%%%%%%%%%%%%%%%%%%%%%%%%%%%%%%%%%%%%%%%%%%%%%%%%%%%%%%%%%%%%%%%%%
\subsection{Proof of Lemma~\ref{lem:combined_Schulz_update}}
\begin{proof}
%%%%%%%%%%%%%%%%%%%%%%%%%%%%%%%%%%%%%%%%%%%%%%%%%%%
We aim to show that $f(\boldsymbol{U},\boldsymbol{W}^+,\boldsymbol{V}) \leq f(\boldsymbol{U},\boldsymbol{W},\boldsymbol{V})$.
This is equivalent to showing
\begin{equation}\label{eq:cost_function_for_Schulz}
    \!\sum_{k=1}^{K} {\!\alpha_k   \{\mathtt{Tr}\big(\boldsymbol{E}_k\boldsymbol{W}_k (\boldsymbol{I} \! - \!\boldsymbol{E}_k \boldsymbol{W}_k)\big)} - \log \det (2\boldsymbol{I} -\boldsymbol{E}_k \boldsymbol{W}_k)\}  \!\leq  \!0 .
\end{equation}
Restricting the attention to the $k^{th}$ user and re-writing the trace and determinant in terms of eigenvalues yields
\begin{equation}\label{eq:cost_function_for_Schulz_one_user}
    \sum_{i=1}^{d}{\!\lambda_i(\boldsymbol{E}_k\boldsymbol{W}_k)} - \sum_{i=1}^{d}{\!\lambda_i^2(\boldsymbol{E}_k\boldsymbol{W}_k)} \leq \sum_{i=1}^{d}{\!\log\lambda_i(2\boldsymbol{I}-\boldsymbol{E}_k\boldsymbol{W}_k)},
\end{equation}
where \mbox{$\lambda_i(\boldsymbol{E}_k\boldsymbol{W}_k\boldsymbol{E}_k\boldsymbol{W}_k)=\lambda^2_i\boldsymbol{(\boldsymbol{E}}_k\boldsymbol{W}_k)$.}
Consider now only the $i^{th}$ term in each summation in~(\ref{eq:cost_function_for_Schulz_one_user}). We can conclude that ${\lambda_i(\boldsymbol{E}_k\boldsymbol{W}_k)} - {\lambda^2_i(\boldsymbol{E}_k\boldsymbol{W}_k)} \leq {\log\lambda_i(2\boldsymbol{I}-\boldsymbol{E}_k\boldsymbol{W}_k)}$ holds whenever $ 0 \leq \lambda_i(\boldsymbol{E}_k\boldsymbol{W}_k) \leq \delta $. Thus, inequality~(\ref{eq:cost_function_for_Schulz_one_user}) holds if \mbox{$\lambda_i(\boldsymbol{E}_k\boldsymbol{W}_k) \leq \delta \; \forall i$} , namely if $\lambda_\mathrm{max}(\boldsymbol{E}_k\boldsymbol{W}_k) \leq \delta $.
We can repeat the same proof for all the users. Hence, we can conclude that inequality~(\ref{eq:cost_function_for_Schulz})
holds if $\lambda_\mathrm{max}(\boldsymbol{E}_k\boldsymbol{W}_k) \leq \delta \; \forall k$.
Note that if $\lambda_\mathrm{max}(\boldsymbol{E}_k\boldsymbol{W}_k) \leq \delta$ then the logarithm is also well-defined.
%%%%%%%%%%%%%%%%%%%%%%%%%%%%%%%%%%%%%%%%%%%%%%%%%%%%%%

From $\boldsymbol{W}_k \succeq 0$ it follows that $\forall i$ 
\begin{equation}\label{eq:lambda_i_2W_minus_WEW}
\lambda_i(2\boldsymbol{W}_k - \boldsymbol{W}_k \boldsymbol{E}_k \boldsymbol{W}_k) = \lambda_i\big((2\boldsymbol{I} - \boldsymbol{W}^{\frac{1}{2}}_k \boldsymbol{E}_k \boldsymbol{W}^{\frac{1}{2}}_k)\boldsymbol{W}^{\frac{1}{2}}_k\boldsymbol{W}^{\frac{1}{2}}_k\big),
\end{equation}
where we have applied the property that the eigenvalues of a product of matrices are invariant under cyclic permutation
of the product order. We observe that $\forall k$ 
\begin{equation}\label{eq:lambda_min_2_minus_WE}
    \lambda_{\mathrm{min}}(2\boldsymbol{I} - \boldsymbol{W}^{\frac{1}{2}}_k \boldsymbol{E}_k \boldsymbol{W}^{\frac{1}{2}}_k) \geq  2 -\lambda_{\mathrm{max}}(\boldsymbol{W}^{\frac{1}{2}}_k \boldsymbol{E}_k \boldsymbol{W}^{\frac{1}{2}}_k) \geq 0,
\end{equation}
where we have applied Weyl's inequality~\cite{Horn1991} and used $\lambda_{\mathrm{max}}(\boldsymbol{E}_k \boldsymbol{W}_k) \leq \delta < 2$. It follows that  
\begin{subequations}
\begin{alignat}{2}
    \lambda_{\mathrm{min}}(\boldsymbol{W}^+_k ) = \lambda_{\mathrm{min}}(2\boldsymbol{W}_k - \boldsymbol{W}_k \boldsymbol{E}_k \boldsymbol{W}_k) &\geq   \\
    \lambda_{\mathrm{min}}(2\boldsymbol{I} - \boldsymbol{W}^{\frac{1}{2}}_k \boldsymbol{E}_k \boldsymbol{W}^{\frac{1}{2}}_k)\lambda_{\mathrm{min}}(\boldsymbol{W}^{\frac{1}{2}}_k\boldsymbol{W}^{\frac{1}{2}}_k) &\geq 0 \; \forall k.
\end{alignat}
\end{subequations}
%%%%%%%%%%%%%%%%%%%%%%%%%%%%%%%%%%%%%%%%%%%%%%%%%%%%%%%%%%%%%

Finally, given that $\boldsymbol{E}_k \succ 0$ and $\boldsymbol{W}_k \succeq 0$, there exists an invertible matrix $\boldsymbol{S}_k \; \forall k$ such that $\boldsymbol{E}_k \boldsymbol{W}_k = \boldsymbol{S}_k^{-1} \boldsymbol{\Lambda}_k \boldsymbol{S}_k$
where $\boldsymbol{\Lambda}_k$ is a diagonal matrix containing the eigenvalues of $\boldsymbol{E}_k \boldsymbol{W}_k$. 
Therefore, we can note that
\begin{align}
\boldsymbol{E}_k \boldsymbol{W}_k^+ =& \boldsymbol{E}_k \boldsymbol{W}_k(2\boldsymbol{I} - \boldsymbol{E}_k \boldsymbol{W}_k) \\
=& \boldsymbol{S}_k^{-1} \boldsymbol{\Lambda}_k \boldsymbol{S}_k(2 \boldsymbol{I} - \boldsymbol{S}^{-1} \boldsymbol{\Lambda} \boldsymbol{S}_k) \\
=& \boldsymbol{S}_k^{-1} \boldsymbol{\Lambda} \boldsymbol{S}_k(2 \boldsymbol{S}_k^{-1} \boldsymbol{S}_k - \boldsymbol{S}_k^{-1} \boldsymbol{\Lambda}_k \boldsymbol{S}_k) \\
=& \boldsymbol{S}_k^{-1} \boldsymbol{\Lambda}(2 \boldsymbol{I}-\boldsymbol{\Lambda}_k) \boldsymbol{S}_k.
\end{align} 
In other words, $\boldsymbol{E}_k \boldsymbol{W}_k^+$ and $\boldsymbol{E}_k \boldsymbol{W}_k$ share the same set of eigenvectors, and the Schulz iteration only updates the eigenvalues of $\boldsymbol{E}_k \boldsymbol{W}_k$ according to
$$
\lambda_i(\boldsymbol{E}_k \boldsymbol{W}_k) \rightarrow \lambda_i(\boldsymbol{E}_k \boldsymbol{W}_k)^+ \triangleq \lambda_i(\boldsymbol{E}_k \boldsymbol{W}_k)(2-\lambda_i(\boldsymbol{E}_k \boldsymbol{W}_k)).
$$
Consequently, as long as $\lambda_{\mathrm{max}}(\boldsymbol{E}_k \boldsymbol{W}_k) \leq \delta < 2$ it follows that $\lambda_{\mathrm{max}}(\boldsymbol{E}_k \boldsymbol{W}^+_k) \leq 1$, namely a single Schulz iteration brings the eigenvalues of $\boldsymbol{E}_k \boldsymbol{W}_k^+$ into the range $[0,1]$.
\end{proof}
%%%%%%%%%%%%%%%%%%%%%%%%%%%%%%%%%%%%%%%%%%%%%%%%%%%%%%%%%%%%%%%%%
\subsection{Proof of Lemma~\ref{lem:step_size_bound_GD_UV}}
%%%%%%%%%%%%%%%%%%%%%%%%%%%%%%%%%%%%%%%%%%%%%%%%%%%%%%%%%%%%%%%%%%%
Note that the reciprocal scaling of $\boldsymbol{U}$ and $\boldsymbol{V}$ does not affect $\boldsymbol{E}$ and consequently does not affect cost function $f$ either. Thus, for the purpose of this proof and for simplicity of notation we do not consider the scaling. We recall that $\lambda_{\mathrm{max}}(\boldsymbol{W}_k\boldsymbol{E}_k) \leq 1 \; \forall k$ is satisfied after $J_w$ Schulz iterations on $\boldsymbol{W}$, as pointed out by Lemma~\ref{lem:combined_Schulz_update}, and it is also satisfied when entering the matrix-inverse-free WMMSE algorithm for the first time because it is satisfied by the initial condition, given by $\boldsymbol{U} = \boldsymbol{0}$ and $\boldsymbol{W} = \boldsymbol{I}$.

We define $\boldsymbol{E}_k^{v^{i}}$ as $\boldsymbol{E}_k$ after taking the $i^{th}$ GD step on $\boldsymbol{V}$, and $\boldsymbol{E}_k^{u^{j}}$ as $\boldsymbol{E}_k$ evaluated after $J_v$ GD steps on $\boldsymbol{V}$ and after taking the $j^{th}$ GD step on $\boldsymbol{U}$. 

Noting that
$\boldsymbol{E}^{+}_k - \boldsymbol{E}_k= \sum_{i=0}^{J_v-1}{(\boldsymbol{E}_k^{v^{i+1}}- \boldsymbol{E}_k^{v^{i}})} + \sum_{j=0}^{J_u-1}{(\boldsymbol{E}_k^{u^{j+1}}- \boldsymbol{E}_k^{u^{j}})} $
leads to
\begin{multline}\label{eq:lambda_max_sum}
    \lambda_{\mathrm{max}}(\boldsymbol{W}_k(\boldsymbol{E}_k^{+}-\boldsymbol{E}_k)) \leq       \sum_{i=0}^{J_v-1}{ \lambda_{\mathrm{max}}\big(\boldsymbol{W}_k{(\boldsymbol{E}_k^{v^{i+1}} - \boldsymbol{E}_k^{v^{i}}})\big)} \\ +
    \sum_{j=0}^{J_u-1}{ \lambda_{\mathrm{max}}\big(\boldsymbol{W}_k(\boldsymbol{E}_k^{u^{j+1}}- \boldsymbol{E}_k^{u^{j}})\big)},
\end{multline}
where we have applied Weyl's inequality~\cite{Horn1991} recalling that $\lambda_{\mathrm{max}}(\boldsymbol{W}_k\boldsymbol{E}_k) = \lambda_{\mathrm{max}}(\boldsymbol{W}^{\frac{1}{2}}_k \boldsymbol{E}_k \boldsymbol{W}^{\frac{1}{2}}_k)$. From Lemma~\ref{lem:combined_Schulz_update} we have that $\lambda_{\mathrm{max}}(\boldsymbol{W}_k\boldsymbol{E}_k) \leq 1 \, \forall k$ and this yields
\begin{equation}\label{eq:lambda_max_WEplus}
\lambda_{\mathrm{max}}(\boldsymbol{W}_k\boldsymbol{E}_k^+) \leq \lambda_{\mathrm{max}}(\boldsymbol{W}_k(\boldsymbol{E}_k^+ - \boldsymbol{E}_k)) + 1 \; \forall k.
\end{equation}
If we assume that $\lambda_{\mathrm{max}}(\boldsymbol{W}_k(\boldsymbol{E}_k^{v^{i+1}} - \boldsymbol{E}_k^{v^i})) \leq \frac{\delta -1}{J_u + J_v}$ for $ i=0,\cdots, J_v-1$ and $\lambda_{\mathrm{max}}(\boldsymbol{W}_k(\boldsymbol{E}_k^{u^{j+1}} - \boldsymbol{E}_k^{u^j})) \leq \frac{\delta -1}{J_u + J_v}$ for $j=0,\cdots, J_u-1$, then combining~(\ref{eq:lambda_max_WEplus}) and~(\ref{eq:lambda_max_sum}) would imply that $\lambda_{\mathrm{max}}(\boldsymbol{W}_k\boldsymbol{E}_k^+) \leq \delta \; \forall k$.
%%%%%%%%%%%%%%%%%%%%%%%%%%%%%%%%%%%%%%%%%%%%%%%%%%%%%%%%%%%%%%%%%%%%%%%
%%%%%%%%%%%%%%%%%%%%%%%%%%%%%%%%%%%%%%%%%%%%%%%%%%%%%%%%%%%%%%%%%%%%%%

We have that $\boldsymbol{E}_k^{v^{i+1}} - \boldsymbol{E}_k^{v^i} = \gamma_v(\boldsymbol{A}_k\gamma_v + \boldsymbol{B}_k)$
where 
\begin{equation}\label{eq:A_def}
\boldsymbol{A}_k = \boldsymbol{U}_k^H\boldsymbol{\Psi}_k\boldsymbol{U}_k,
\end{equation}
\begin{equation}\label{eq:B_def}
\boldsymbol{B}_k = \nabla_v f^H(\boldsymbol{V}_k) \boldsymbol{H}_k^H \boldsymbol{U}_k + {\boldsymbol{U}_k}^H \boldsymbol{H}_k \nabla_v f(\boldsymbol{V}_k) - {\boldsymbol{U}_k}^H\boldsymbol{\Omega}\boldsymbol{U}_k,
\end{equation} 
where $\boldsymbol{\Psi}$ and $\boldsymbol{\Omega}$ are given in~(\ref{eq:def_Psi}) and in~(\ref{eq:def_Omega}), respectively.
It follows that $\lVert \boldsymbol{W}_k^{\frac{1}{2}} \boldsymbol{A}_k\boldsymbol{W}_k^{\frac{1}{2}}  \rVert_2 \leq L_a$
and
$\lVert \boldsymbol{W}_k^{\frac{1}{2}} \boldsymbol{B}_k\boldsymbol{W}_k^{\frac{1}{2}}  \rVert_2 \leq L_b$  $\forall k$,
where $L_a$ and $L_b$ are given in~(\ref{eq:def_nu_v}) and $\mu_v$ therein is an upper bound on $\lVert \nabla_v f\rVert_2$.
Thus, for $ i=0,\cdots, J_v-1$ we have that
\begin{equation}
\lVert\boldsymbol{W}_k^{\frac{1}{2}}(\boldsymbol{E}_k^{v^{i+1}} - \boldsymbol{E}^{v^{i}}_k) \boldsymbol{W}_k^{\frac{1}{2}}   \rVert_2 \leq
L_a\gamma^2_v + L_b\gamma_v \; \forall k.
\end{equation}
This implies that for $ i=0,\cdots, J_v-1$ and $ 0 < \gamma_v \leq \nu_v$
\begin{equation}
\lambda_{\mathrm{max}}(\boldsymbol{W}_k(\boldsymbol{E}_k^{v^{i+1}} - \boldsymbol{E}_k^{v^i})) \leq \frac{\delta -1}{J_u + J_v}.
\end{equation}

%%%%%%%%%%%%%%%%%%%%%%%%%%%%%%%%%%%%%%%%%%%%%%%%%%%
For $\boldsymbol{U}$ we have that 
$\boldsymbol{E}_k^{u^{j+1}} - \boldsymbol{E}_k^{u^j} = \gamma_u(\boldsymbol{C}_k\gamma_u + \boldsymbol{D}_k)$,
where 
\begin{equation}\label{eq:C_def}
    \boldsymbol{C}_k =
   \nabla_u {f}^H (\boldsymbol{U}_k)  \Big(\boldsymbol{\Upsilon}_k + \boldsymbol{I}\sigma^2\Big) \nabla_u  f(\boldsymbol{U}_k),
\end{equation}
\begin{align}\label{eq:D_def}
    &\boldsymbol{D}_k = {\boldsymbol{V}_k}^H\boldsymbol{H}_k^H \nabla f_u (\boldsymbol{U}_k) + \nabla_u {f}^H ({\boldsymbol{U}}_k) \boldsymbol{H}_k \boldsymbol{V}_k \nonumber\\- {\boldsymbol{U}}_k^H \big(&\boldsymbol{\Upsilon}_k + \boldsymbol{I}\sigma^2 \big) \nabla f_u (\boldsymbol{U}_k) - \nabla_u {f}^H (\boldsymbol{U}_k) \big(\boldsymbol{\Upsilon}_k + \boldsymbol{I}\sigma^2 \big)\boldsymbol{U}_k,
\end{align}
where $\boldsymbol{\Upsilon}_k = \sum_{m=1}^{K}{\boldsymbol{H}_k\boldsymbol{V}_m{\boldsymbol{V}_m}^H\boldsymbol{H}^H_k}$.
It follows that $\lVert \boldsymbol{W}_k^{\frac{1}{2}} \boldsymbol{C}_k\boldsymbol{W}_k^{\frac{1}{2}}  \rVert_2 \!\leq \!L_c$ and $\lVert \boldsymbol{W}_k^{\frac{1}{2}} \boldsymbol{D}_k\boldsymbol{W}_k^{\frac{1}{2}}  \rVert_2 \!\leq \!L_d$ $\forall k$,
where $L_c$ and $L_d$ are given in~(\ref{eq:def_nu_u}) and $\mu_u$ therein is an upper bound on $\lVert \nabla_u f\rVert_2$.
%%%%%%%%%%%%%%%%%%%%%%%%%%%%%%%%%%%%%%%%%%%%%%%%%%%
Thus, for $ j=0,\cdots, J_u-1$
\begin{equation}
\lVert\boldsymbol{W}_k^{\frac{1}{2}}(\boldsymbol{E}^{u^{j+1}}_k - \boldsymbol{E}^{u^{j}}_k) \boldsymbol{W}_k^{\frac{1}{2}}   \rVert_2 \leq 
L_c\gamma^2_u + L_d\gamma_u \; \forall k.
\end{equation}
This implies that for $ j=0,\cdots, J_u-1$ and $0 < \gamma_u \leq \nu_u$
\begin{equation}
\lambda_{\mathrm{max}}(\boldsymbol{W}_k(\boldsymbol{E}_k^{u^{j+1}} - \boldsymbol{E}_k^{u^j})) \leq \frac{\delta -1}{J_u + J_v}\raisepunct{.}
\end{equation}
%%%%%%%%%%%%%%%%%%%%%%%%%%%%%%%%%%%%%%%%%%%%%%%%%%%%%%%%%%%%%%%%%%%%%%%
\begin{figure*}
\begin{equation}\label{eq:def_Psi}
\boldsymbol{\Psi}_k = \boldsymbol{H}_k \sum_{m=1}^{K}{\nabla_v f(\boldsymbol{V}_m)}\nabla_v f^H(\boldsymbol{V}_m) \boldsymbol{H}_k^H +  \boldsymbol{I}\frac{\sum_{n=1}^{K}{\mathtt{Tr}\big(\nabla_v f(\boldsymbol{V}_n) \nabla_v f^H(\boldsymbol{V}_n)\big)}\sigma^2}{P}
\end{equation}
\begin{equation}\label{eq:def_Omega}
    \boldsymbol{\Omega}_k = \boldsymbol{H}_k \sum_{m=1}^{K}{\nabla_v f(\boldsymbol{V}_m)  \boldsymbol{V}^H_m} \boldsymbol{H}_k^H + \boldsymbol{H}_k  \sum_{m=1}^{K}{ \boldsymbol{V}_m \nabla_v f^H(\boldsymbol{V}_m)} \boldsymbol{H}_k^H  +\boldsymbol{I}\frac{\sum_{n=1}^{K}{\mathtt{Tr}\Big(\boldsymbol{V}_n \nabla_v f^H(\boldsymbol{V}_n) + \nabla_v f(\boldsymbol{V}_n) \boldsymbol{V}^H_n}\Big)\sigma^2}{P}
\end{equation}
\end{figure*}

Thus, by using~(\ref{eq:lambda_max_WEplus}) we can conclude that for $0 < \gamma_u \leq \nu_u$ and $0 < \gamma_v \leq \nu_v$, $\lambda_{\mathrm{max}}(\boldsymbol{W}_k\boldsymbol{E}_k^+) \leq \delta \; \forall k$ holds. 

%%%%%%%%%%%%%%%%%%%%%%%%%%%%%%%%%%%%%%%%%%%%%%%%%%%%%%%%%%%%%%%%%%%
%%%%%%%%%%%%%%%%%%%%%%%%%%%%%%%%%%%%%%%%%%%%%%%%%%%%%%%%%%%%%%%%%%%
%%%%%%%%%%%%%%%%%%%%%%%%%%%%%%%%%%%%%%%%%%%%%%%%%%%%%%%%%%%%%%%%%%%

%%%%%%%%%%%%%%%%%%%%%%%%%%%%%%%%%%%%%%%%%%%%%%%%%%%%%%%%%%%%%%%%%%%%
\subsection{Proof of Lemma~\ref{lem:limit_point}}
\begin{proof}
Applying Theorem~\ref{theo:monotonicity} and setting \mbox{$\boldsymbol{U}= \boldsymbol{0}$} and \mbox{$\boldsymbol{W} = \boldsymbol{I}$} as initialization leads to 
$
d \geq \mathtt{Tr}(\boldsymbol{W}^{l}_k\boldsymbol{E}^{l}_k) - \log\det (\boldsymbol{W}^{l}_k)$ $\forall l,k$,
where $l$ is the algorithm iteration index. By noting that the optimal update of $\boldsymbol{W}^{l}_k$ is given by the inverse of $\boldsymbol{E}^{l}_k$, it follows that
$d \geq d + \log \det (\boldsymbol{E}^{l}_k)$ $\forall l,k$.
This implies that  
$\log \det \boldsymbol{E}^l_k = \sum_{i=1}^{d}{\log \lambda_i(\boldsymbol{E}^l_k)} \leq 0 \; \; \forall l,k,
$ and by Lemma~\ref{lem:lower_bound_E} we have \mbox{$\lambda_{\mathrm{max}}(\boldsymbol{E}_k) \leq \! \big({\frac{P\kappa + \sigma^2}{\sigma^2}}\big)^{d-1}$} $\forall k$. It follows that $\forall k $
\begin{equation}\label{eq:U_confined_set}
\lVert \boldsymbol{U}_k\rVert ^2_F = \mathtt{Tr}(\boldsymbol{U}_k^H\boldsymbol{U}_k) \leq \frac{\mathtt{Tr}(\boldsymbol{E}_k)}{\sigma^2} \leq   \frac{d}{\sigma^2}{\Big(\frac{P\kappa + \sigma^2}{\sigma^2} \Big)}^{d-1}
\end{equation}
where $\sum_{m=1}^{K}{\mathtt{Tr}(\boldsymbol{V}_m \boldsymbol{V}^H_m)}=P$.
Combining~(\ref{eq:condition_eig_EW_delta}) and Lemma~\ref{lem:lower_bound_E} yields $\lambda_{\mathrm{max}}(\boldsymbol{W}_k) \leq \delta \frac{P\kappa + \sigma^2}{\sigma^2}\; \forall k,$ thus we have
\begin{equation}\label{eq:W_confined_set}
    \lVert \boldsymbol{W}_k \rVert^2_F \leq d \delta^2 \frac{(P\kappa + \sigma^2)^2}{\sigma^4}\; \; \forall k  
\end{equation}
and from the scaling operation on $\boldsymbol{V}$ we have 
\begin{equation}\label{eq:V_confined_set}
    \lVert \boldsymbol{V}_k\rVert^2_F \leq P \; \; \forall k. 
\end{equation}

Combining~(\ref{eq:U_confined_set}), (\ref{eq:W_confined_set}), and~(\ref{eq:V_confined_set}) implies that the sequence $\{(\boldsymbol{U},\boldsymbol{W},\boldsymbol{V})^l\}$ is confined to a compact set and by Bolzano-Weierstrass theorem~\cite{bartle2011} it follows that there exists a subsequence $\{(\boldsymbol{U},\boldsymbol{W},\boldsymbol{V})^{l_j}\}$ converging to a limit point.
\end{proof}

%%%%%%%%%%%%%%%%%%%%%%%%%%%%%%%%%%%%%%%%%%%%%%%%%%%%%%%%%%%%%%%%%%%
\subsection{Proof of Lemma~\ref{lem:stationary_points_equivalence}}
%%%%%%%%%%%%%%%%%%%%%%%%%%%%%%%%%%%%%%%%%%%%%%%%%%%%%%%%%%%%%%%%%%%
\begin{proof}
We define 
\begin{equation}\nonumber
\tilde{f}(\boldsymbol{U},\boldsymbol{W},\boldsymbol{V}) \triangleq \sum_{k = 1}^{K}\alpha_k \big\{  \mathtt{Tr}(\boldsymbol{W}_k{\widetilde{\boldsymbol{E}}}_{k}) - \log\det(\boldsymbol{W}_k)\big\}
\end{equation}
as the objective function in~(\ref{eq:equivalent_reformulation}).
Given that $(\overline{\boldsymbol{U}},\overline{\boldsymbol{W}},\overline{\boldsymbol{V}})$ is stationary point of~(\ref{eq:equivalent_reformulation_tilde}), %it follows that
\begin{equation}
\langle \nabla_u f(\overline{\boldsymbol{U}}),\boldsymbol{U}-\overline{\boldsymbol{U}}\rangle  \; \geq 0 \; \forall \boldsymbol{U}
\end{equation}
\begin{equation}
\langle \nabla_v f(\overline{\boldsymbol{V}}),\boldsymbol{V}-\overline{\boldsymbol{V}}\rangle  \; \geq 0 \; \forall \boldsymbol{V},
\end{equation}
where the notation $\langle \cdot,\cdot\rangle $ is given by Definition~\ref{def:inner_product}.

From~(\ref{eq:gradient_u_expression}) and~(\ref{eq:gradient_v_expression}), we note that
\begin{equation}
    \nabla_u f({\overline{\boldsymbol{U}}\frac{1}{\beta},\overline{\boldsymbol{W}},\overline{\boldsymbol{U}}\beta}) = \beta\nabla_u f(\overline{\boldsymbol{U}},\overline{\boldsymbol{W}},\overline{\boldsymbol{U}})
\end{equation}
\begin{equation}
    \nabla_v f({\overline{\boldsymbol{U}}\frac{1}{\beta},\overline{\boldsymbol{W}},\overline{\boldsymbol{V}}\beta})= \frac{1}{\beta}\nabla_v f(\overline{\boldsymbol{U}},\overline{\boldsymbol{W}},\overline{\boldsymbol{V}}).
\end{equation}
This implies that 
\begin{equation}
\langle\nabla_u f({\overline{\boldsymbol{U}}\frac{1}{\beta},\overline{\boldsymbol{W}},\overline{\boldsymbol{V}}\beta}),\boldsymbol{U}-\overline{\boldsymbol{U}}\rangle \; \geq 0 \; \forall \boldsymbol{U}
\end{equation}
\begin{equation}
\langle\nabla_v f({\overline{\boldsymbol{U}}\frac{1}{\beta},\overline{\boldsymbol{W}},\overline{\boldsymbol{V}}\beta}),\boldsymbol{V}-\overline{\boldsymbol{V}}\rangle \; \geq 0 \; \forall \boldsymbol{V}.
\end{equation}
The scaling of $\overline{\boldsymbol{V}}$ ensures that the power constraint is satisfied with equality, therefore we can restrict the feasible set to \mbox{$\mathcal{V}=\{ \boldsymbol{V} | \sum_{m=1}^{K}{\mathtt{Tr}(\boldsymbol{V}_m\boldsymbol{V}_m^H)} \leq  P\}$}.
At $({\overline{\boldsymbol{U}}\frac{1}{\beta},\overline{\boldsymbol{W}},\overline{\boldsymbol{V}}\beta})$ the power constraint is satisfied with equality, thus
\begin{equation}
    \nabla_u \widetilde{f}({\overline{\boldsymbol{U}}\frac{1}{\beta},\overline{\boldsymbol{W}},\overline{\boldsymbol{V}}\beta}) = \nabla_u {f}({\overline{\boldsymbol{U}}\frac{1}{\beta},\overline{\boldsymbol{W}},\overline{\boldsymbol{V}}\beta})
\end{equation}

\begin{equation}
    \nabla_w \widetilde{f}({\overline{\boldsymbol{U}}\frac{1}{\beta},\overline{\boldsymbol{W}},\overline{\boldsymbol{V}}\beta}) = \nabla_w {f}({\overline{\boldsymbol{U}}\frac{1}{\beta},\overline{\boldsymbol{W}},\overline{\boldsymbol{V}}\beta}),
\end{equation}
which implies that
\begin{equation}\label{eq:stationarity_condition_u}
\langle\nabla_u \widetilde{f}({\overline{\boldsymbol{U}}\frac{1}{\beta},\overline{\boldsymbol{W}},\overline{\boldsymbol{V}}\beta}),\boldsymbol{U}-\overline{\boldsymbol{U}}\rangle \; \geq 0 \; \forall \boldsymbol{U}
\end{equation}
\begin{equation}\label{eq:stationarity_condition_w}
\langle\nabla_w \widetilde{f}({\overline{\boldsymbol{U}}\frac{1}{\beta},\overline{\boldsymbol{W}},\overline{\boldsymbol{V}}\beta}),\boldsymbol{W}-\overline{\boldsymbol{W}} \rangle \; \geq 0 \; \forall \, \boldsymbol{W} \in \mathcal{W},
\end{equation}
where $\mathcal{W}$ is the set of positive semidefinite $\boldsymbol{W}$.
Let \mbox{$\tau = \frac{2\sigma^2}{P}\sum_{m=1}^{K}{\alpha_m \mathtt{Tr}(\boldsymbol{W}_m \boldsymbol{U}^H_m \boldsymbol{U}_m)}$}, where $\tau \ge 0$. We have
\begin{equation}
    \nabla_v \widetilde{f}({\overline{\boldsymbol{U}}\frac{1}{\beta},\overline{\boldsymbol{W}},\overline{\boldsymbol{V}}\beta}) = \nabla_v {f}({\overline{\boldsymbol{U}}\frac{1}{\beta},\overline{\boldsymbol{W}},\overline{\boldsymbol{V}}\beta})-2\tau\overline{\boldsymbol{V}}
\end{equation}
and we note that $\langle{\overline{\boldsymbol{V}}},\boldsymbol{V} - \overline{\boldsymbol{V}}\rangle \;=\; \langle{\overline{\boldsymbol{V}}},\boldsymbol{V}\rangle - \lVert \overline{\boldsymbol{V}}\rVert_F^2
$. Given that we restrict the attention only to $\boldsymbol{V}$ such that \mbox{$\lVert \boldsymbol{V} \rVert_F \leq \sqrt{P}$} and that $\lVert \overline{\boldsymbol{V}}\rVert_F =  \sqrt{P}$, it follows that
$\langle{\overline{\boldsymbol{V}}},\boldsymbol{V} - \overline{\boldsymbol{V}}\rangle\leq 0 \; \forall \,\boldsymbol{V} \in \mathcal{V}$. This implies that
\begin{equation}\label{eq:stationarity_condition_v}
    \langle\nabla_v \widetilde{f}({\overline{\boldsymbol{U}}\frac{1}{\beta},\overline{\boldsymbol{W}},\overline{\boldsymbol{V}}\beta}),\boldsymbol{V} - \overline{\boldsymbol{V}}\rangle\; \geq 0 \; \forall \boldsymbol{V} \in \mathcal{V},
\end{equation}
where we have used that \mbox{$ \langle \nabla_v {f}({\overline{\boldsymbol{U}}\frac{1}{\beta},\overline{\boldsymbol{W}},\overline{\boldsymbol{V}}\beta}),\boldsymbol{V} - \overline{\boldsymbol{V}}\rangle \ge 0$}. 
By combining~(\ref{eq:stationarity_condition_u}), (\ref{eq:stationarity_condition_w}), and (\ref{eq:stationarity_condition_v}), we can conclude that $(\overline{\boldsymbol{U}}\frac{1}{\beta},\overline{\boldsymbol{W}},\overline{\boldsymbol{V}}\beta)$ is a stationary point of~(\ref{eq:equivalent_reformulation}).
\end{proof}}
\bibliographystyle{IEEEtran}
\bibliography{bibliography.bib}

\end{document}